\documentclass[acmtog]{acmart}
\AtBeginDocument{%
  }

\setcopyright{acmlicensed}
\copyrightyear{2025}
\acmYear{2025}
\acmDOI{10.1145/3721238.3730620}

\acmConference[SIGGRAPH Conference Papers '25]{Special Interest Group on Computer Graphics and Interactive Techniques Conference Conference Papers }{August 10--14, 2025}{Vancouver, BC, Canada}
\acmBooktitle{Special Interest Group on Computer Graphics and Interactive Techniques Conference Conference Papers (SIGGRAPH Conference Papers '25), August 10--14, 2025, Vancouver, BC, Canada}
\acmISBN{979-8-4007-1540-2/2025/08}

\acmSubmissionID{290}

\usepackage{graphicx}
\usepackage{subcaption}

\usepackage{multirow}

\usepackage{algorithmic}
\usepackage{algorithm}

\newcommand{\argmin}{\mathop{\arg\min}}

\definecolor{green}{rgb}{0.3,0.56,0.0}
\newcommand{\mengyu}[1]{{\color{black}#1}}

\newcommand{\jr}[1]{}

\newcommand{\cmyy}[1]{}
\newcommand{\cmy}[1]{\textcolor{black}{#1}}

\newcommand{\xing}[1]{\textcolor{black}{#1}}
\begin{document}

\title{Gaussian Fluids: A Grid-Free Fluid Solver based on Gaussian Spatial Representation}

\author{Jingrui Xing}
\email{xjr01@hotmail.com}
\orcid{0000-0001-7219-9969}
\affiliation{%
    \institution{School of Intelligence Science and Technology, Peking University}
    \city{Beijing}
    \country{China}
}

\author{Bin Wang}
\email{binwangbuaa@gmail.com}
\orcid{0000-0001-9496-772X}
\affiliation{
    \institution{Independent Researcher}
    \city{Beijing}
    \country{China}
}

\author{Mengyu Chu}
\email{mchu@pku.edu.cn}
\affiliation{
    \institution{State Key Laboratory of General Artificial Intelligence, Peking University}
    \city{Beijing}
    \country{China}
}
\authornote{Corresponding authors}

\author{Baoquan Chen}
\email{baoquan@pku.edu.cn}
\orcid{0000-0003-4702-036X}
\affiliation{
    \institution{State Key Laboratory of General Artificial Intelligence, Peking University}
    \city{Beijing}
    \country{China}
}
\authornotemark[1]


\begin{abstract}
We present a grid-free fluid solver featuring a novel Gaussian representation.
Drawing inspiration from the expressive capabilities of 3D Gaussian Splatting in multi-view image reconstruction, we model the continuous flow velocity as a weighted sum of multiple Gaussian functions. 
\cmy{This representation is continuously differentiable, which enables us to derive spatial differentials directly and solve the time-dependent PDE via a custom first‑order optimization tailored to fluid dynamics.}
\cmy{Compared to traditional discretizations, which typically adopt Eulerian, Lagrangian, or hybrid perspectives, our approach is inherently memory-efficient and spatially adaptive, enabling it to preserve fine-scale structures and vortices with high fidelity. While \xing{these advantages} are also sought by implicit neural representations, GSR offers enhanced robustness, accuracy, and generality across diverse fluid phenomena, with improved computational efficiency during temporal evolution.}
\cmy{Though our first‑order solver does not yet match} 
the \cmy{speed} of fluid solvers \cmy{using explicit representations}, \cmy{its continuous nature substantially reduces spatial discretization error and opens a new avenue for high‑fidelity simulation.}
We evaluate the proposed solver across a broad range of 2D and 3D fluid phenomena,
\cmy{demonstrating its ability to preserve intricate vortex dynamics, accurately capture boundary-induced effects such as Kármán vortex streets, and remain robust across long time horizons—all without additional parameter tuning. Our results suggest that GSR offers a compelling direction for future research in fluid simulation.}
The source code for our fluid solver is publicly available at \url{https://github.com/xjr01/Gaussian-Fluids-Code}.
\end{abstract}


\begin{teaserfigure}
    \centering
    \newcommand{\formattedgraphics}[1]{\includegraphics[trim=700 300 1200 250,clip,width=0.195\textwidth]{#1}}
    \formattedgraphics{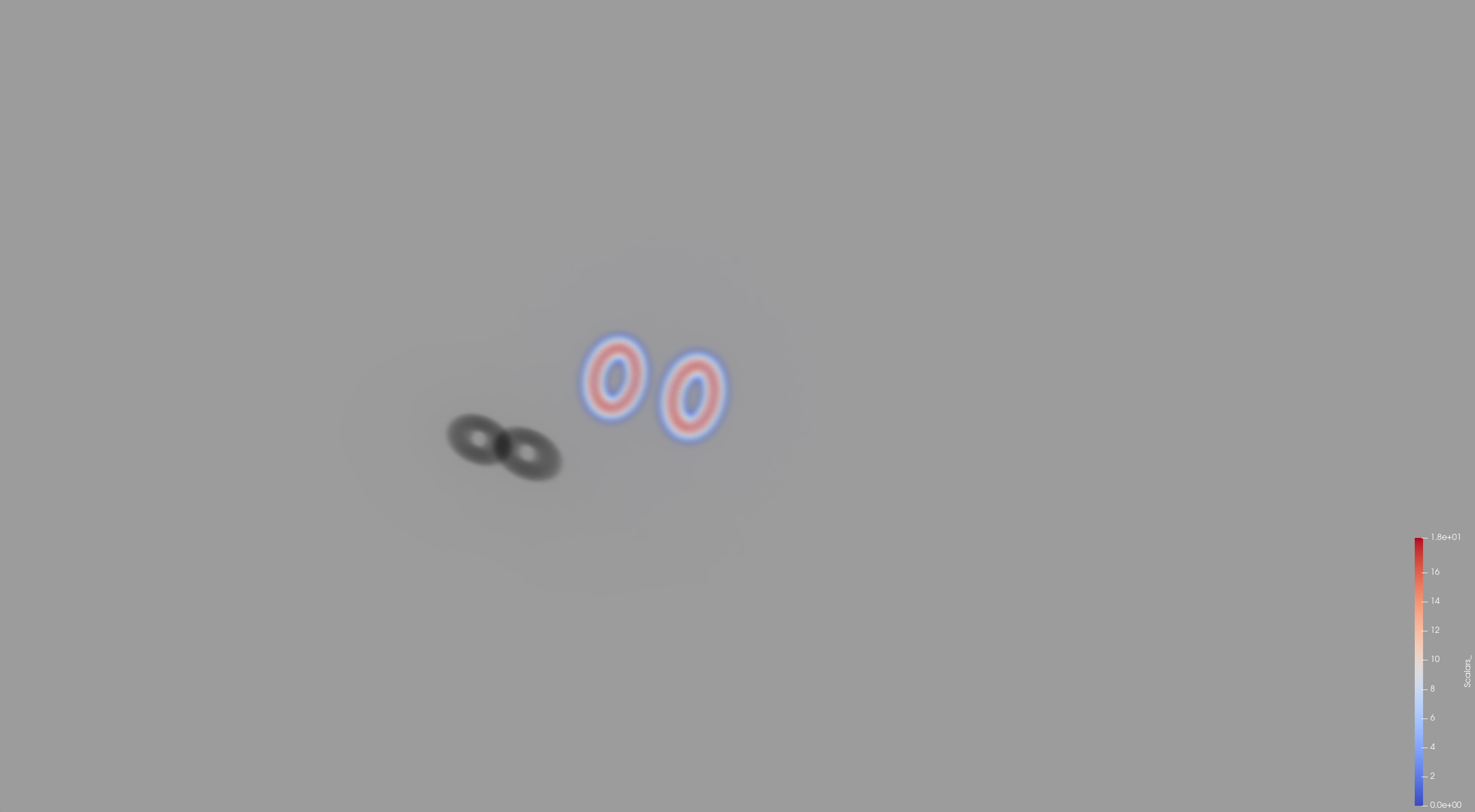}
    \hspace{-4.7pt}
    \formattedgraphics{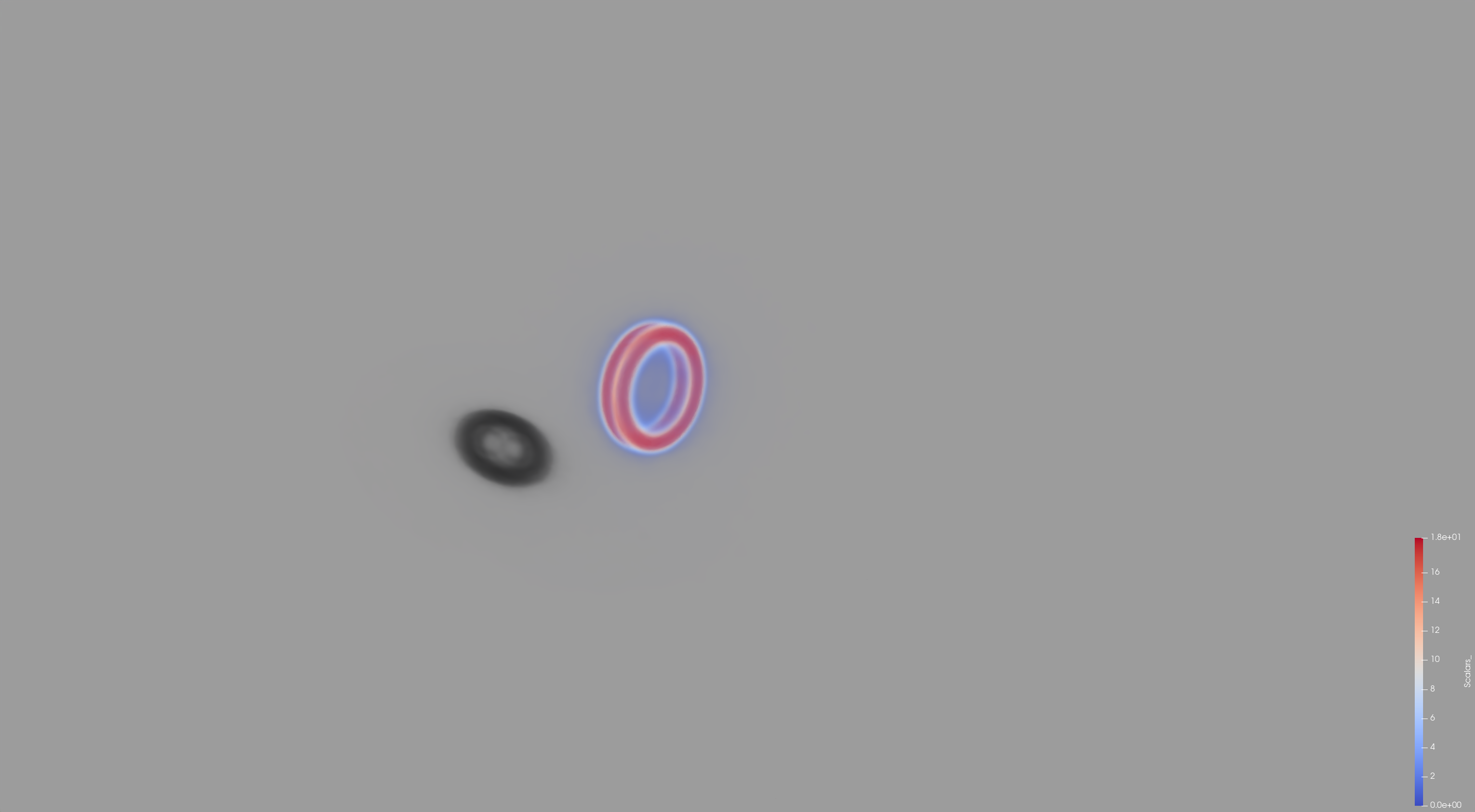}
    \hspace{-4.7pt}
    \formattedgraphics{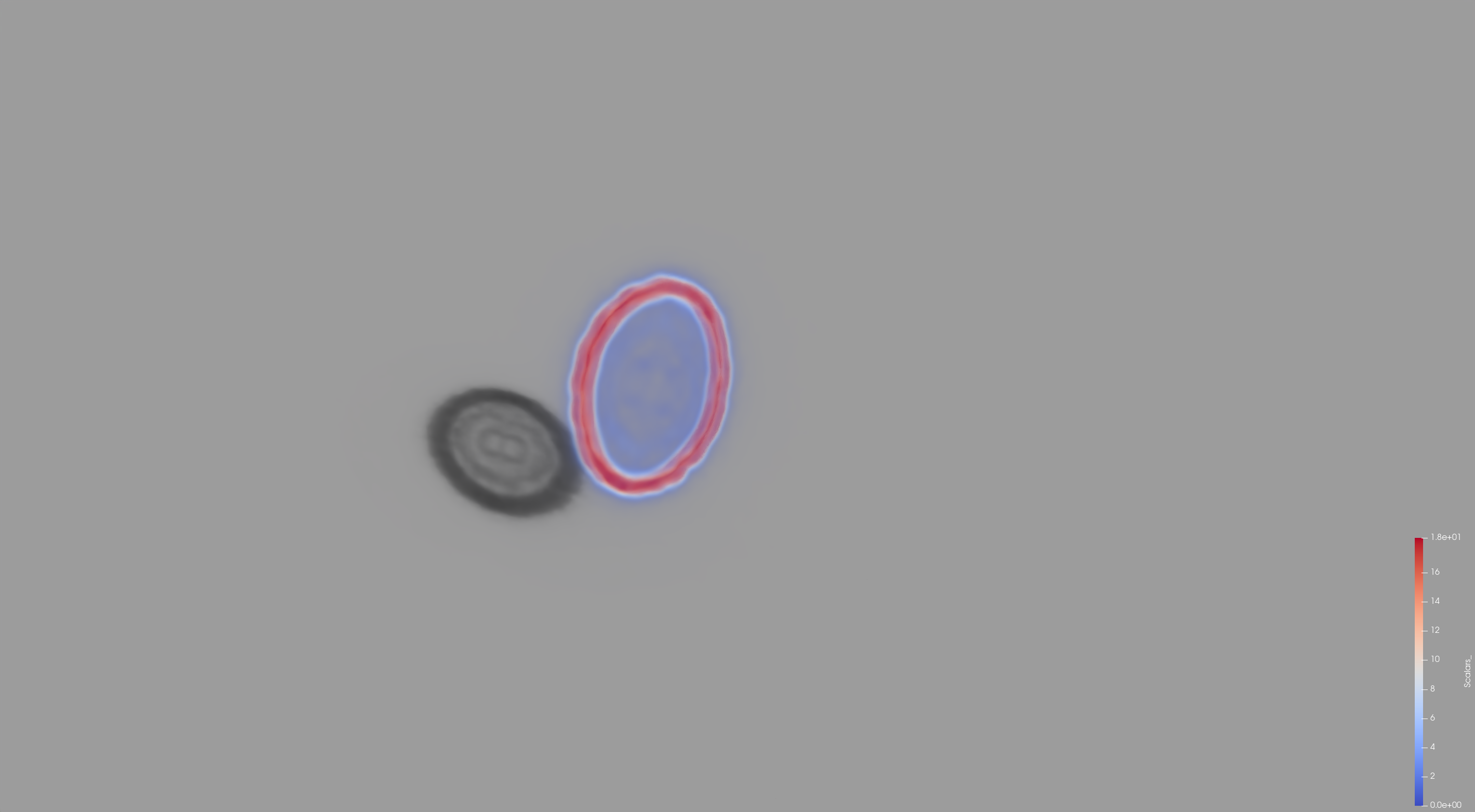}
    \hspace{-4.7pt}
    \formattedgraphics{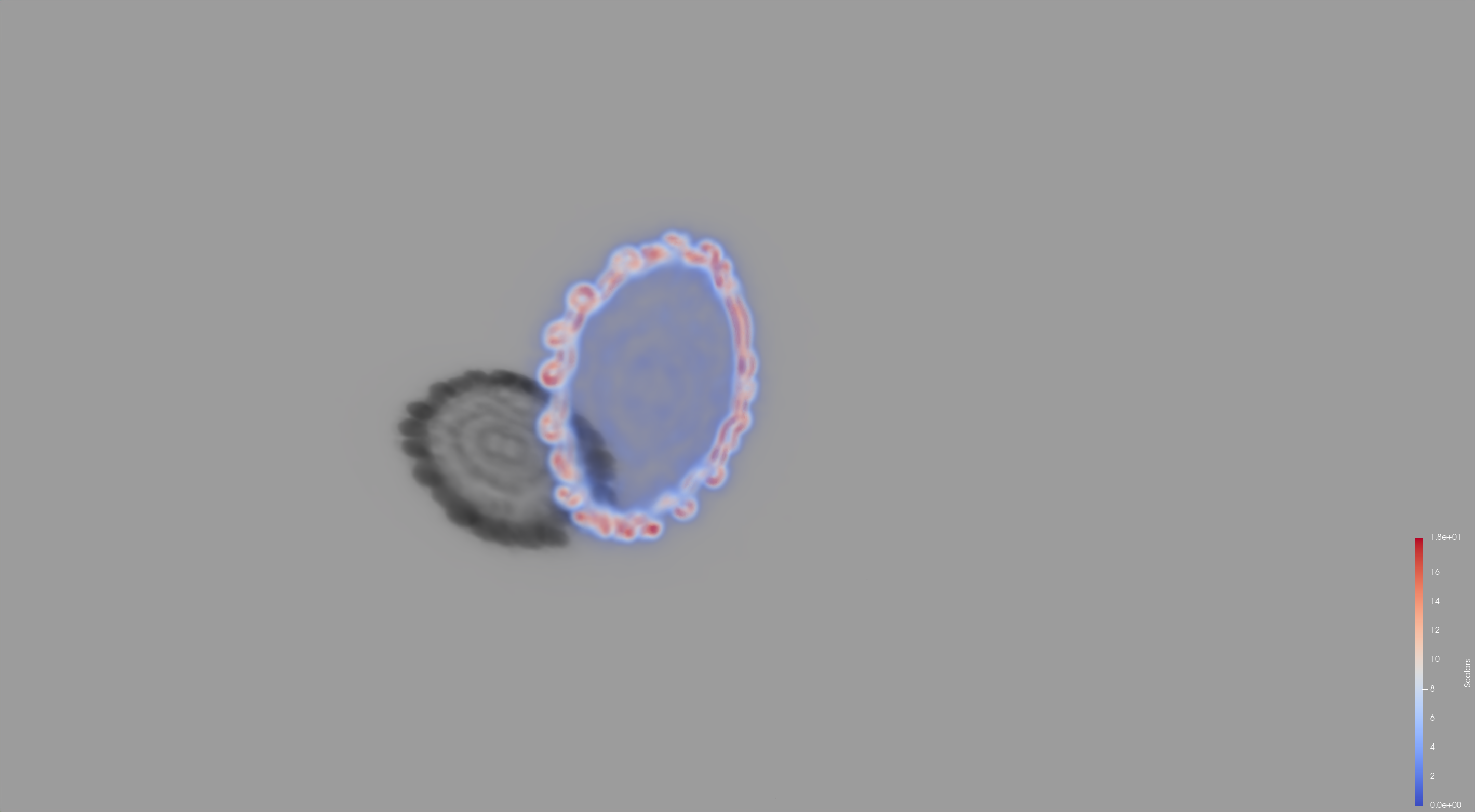}
    \hspace{-4.7pt}
    \includegraphics[trim=700 300 1100 250,clip,width=0.2152\textwidth]{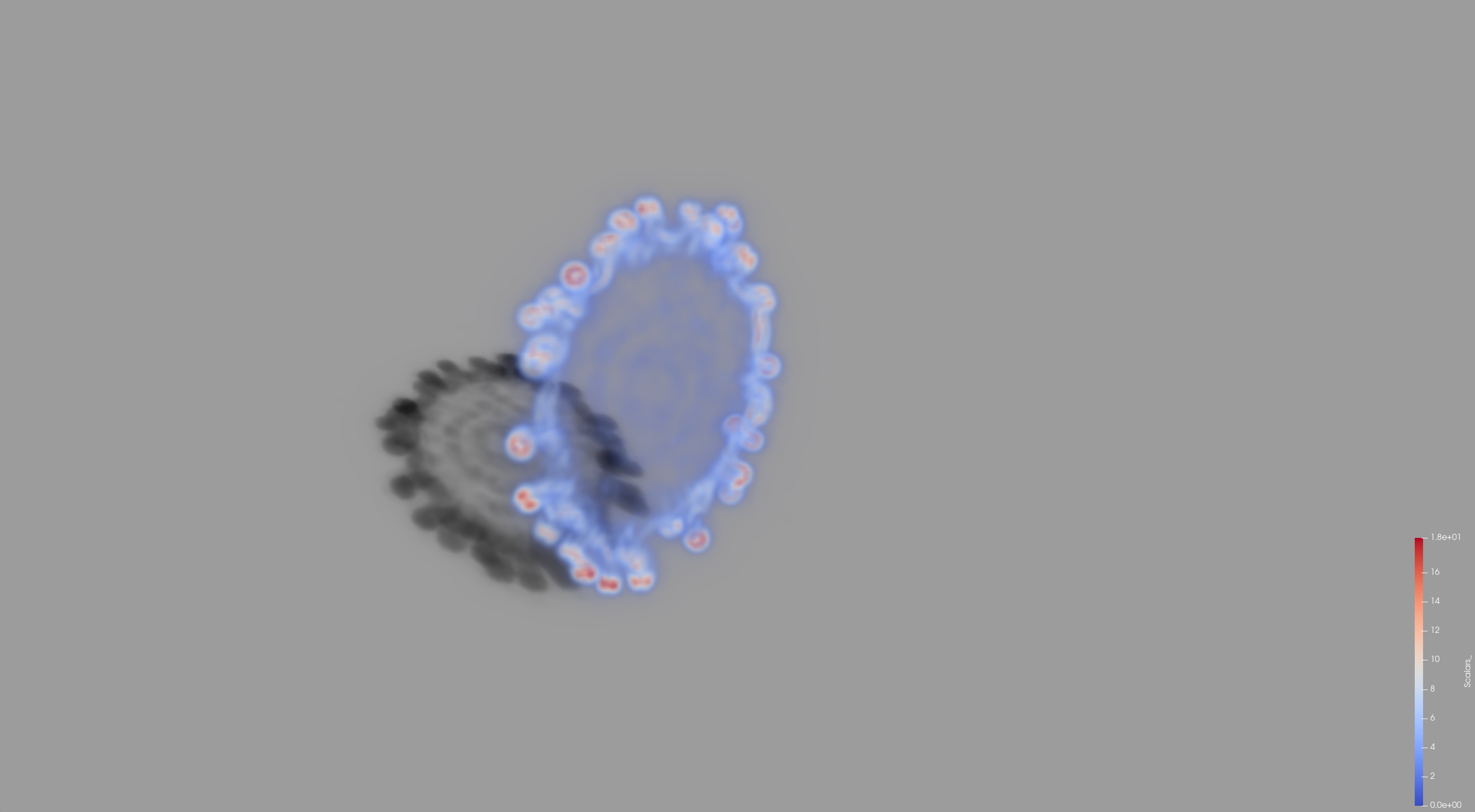}
    \vspace{-8pt}
    \caption{Simulation of two vortex rings colliding using our grid-free fluid solver based on Gaussian Spatial Representation. The figures visualize the vorticity magnitude at frames 0, 74, 123, 178, and 242, showcasing long temporal stability, improved vorticity preservation, and adaptive spatial accuracy, capturing fine-scale details with intricate dynamics, such as the disturbed large vortex ring splitting into massive smaller ones.}
    \label{fig:ring_collide}
\end{teaserfigure}


\maketitle

\section{Introduction}
Fluid phenomena have presented enduring challenges in computer graphics, characterized by the need for expressive representations to capture rich spatial details, alongside the demand for efficient and accurate temporal evolution to handle non-linear and chaotic dynamics. 
Traditional fluid solvers lean on grid-based or particle-based spatial representations, aligning with the Eulerian and Lagrangian perspectives. Hybrid approaches have also been proposed.
These straightforward spatial discretizations enable efficient calculations for solving temporal dynamics.

\cmy{Despite substantial progress, }traditional solvers persistently face challenges due to limited expressiveness. Eulerian methods \cmy{often suffer from} numerical viscosity arising from the lack of continuity, while the limited accuracy of \xing{Lagrangian methods like the} Smoothed Particle Hydrodynamics (SPH) impedes the capture of fine structures\cmy{~\cite{KBST2022}}.
Hybrid \cmy{approaches aim to combine the strengths of both views, }\xing{but introduce numerical error in data transferring between the two discretizations.}
\xing{Though these limitations can be alleviated with extensive memory usage, it would lead to the curse of dimensionality.}

\xing{To deal with these challenges, }we aim to utilize expressive\cmy{, continuous} spatial representations for fluid simulation.
In this context, both Implicit Neural Representations (INR) and 3D Gaussian functions 
\cmy{have recently shown promise in representing both function values and derivatives with high fidelity and spatial adaptiveness. }
While we focus on Gaussian functions due to their flexibility and efficiency,
our endeavor aligns with the growing trend of employing INR in simulations, 
e.g. for fluid dynamics~\cite{chen2023implicit}, garment untangling~\cite{santesteban2022ulnef}, and facial animation~\cite{yang2023implicit}.
\cmy{Yet, the continuous nature of these representations often incurs greater computational cost and optimization challenges, such as slow convergence~\cite{wang2022and} and difficulty in enforcing hard physical constraints~\cite{chen2023implicit}.
There remains no consensus on an efficient, unified PDE‑solving strategy or interactive editing workflow. }

\cmy{We propose Gaussian Spatial Representation (GSR),  a novel continuous representation for fluid dynamics, paired with a first‑order, physics‑guided optimization framework. We carefully design a composite loss that enforces incompressibility and preserves vorticity. We also introduce an optimization strategy to resolve conflicts in gradient directions, enhancing stability and convergence. Our solution consistently outperforms INR‑based approaches in stability, accuracy, and detail preservation.}

\cmy{In summary, GSR offers the following properties:}
\begin{itemize}
    \item \cmy{Continuous Differentiability: Enables accurate and efficient computation of spatial derivatives for PDE terms.}
    \item \cmy{Compactness and Adaptivity: Significantly reduces memory footprint while retaining thin structures.}
    \item \cmy{Vorticity Fidelity: Maintains vortex structures while handles harmonic components more faithfully than vortex‑only approaches.}
\end{itemize}
\cmy{We acknowledge that the first‑order solver speed and strict constraint enforcement still lag behind well-established discretizations, but our results, achieved without per‑scene parameter tuning, demonstrate a stable, spatially adaptive solver. It performs well across various 2D and 3D scenarios, with and without obstacles, charting a promising direction for continuous, grid‑free fluid simulation.}

\section{Related Work}

\paragraph{Traditional Fluid Simulation Methods}
We briefly summarize traditional simulation methods with a focus on single-phase flows. 
Since the introduction of the stable fluids algorithm\cmy{~\cite{Stam99}}, various methods have been developed to solve the highly nonlinear Navier-Stokes equations following the operator-splitting technique.
Grid-based methods primarily focus on accelerating the pressure projection step~\cite{McAdams10, Aanjaneya17} and reducing numerical dissipation with more accurate advection~\cite{kim2005flowfixer, selle2008unconditionally}. \cmy{Recently, there has been growing interest in impulse-based fluid methods~\cite{Feng2023Impulse, nabizadeh2022covector, selle2008unconditionally}, which refine advection through a reformulation of the governing equations into the impulse-velocity framework. These methods lead to better vorticity preservation, especially within Eulerian velocity-based solvers, and demonstrate improved stability in capturing fine-scale features.}

Lagrangian alternatives such as Smoothed-particle hydrodynamics (SPH) methods~\cite{muller2003particle} have been widely explored. These methods focus on enhancing incompressibility ~\cite{ihmsen2013implicit, bender2016divergence}, \cmy{improving spatial adaptivity~\cite{Owen_1998}}, and mitigating errors due to interpolation inconsistencies in sparsely sampled regions~\cite{band2018mls, westhofen2023comparison}.
\cmy{A comprehensive overview of SPH techniques can be found in~\cite{Koschier2019SPH}.}
Vortex methods typically use Lagrangian representations such as particles~\cite{cottet2000vortex}, filaments~\cite{weissmann2010filament}, and sheets~\cite{brochu2012linear}.
They reformulate the fluid equations into vorticity-velocity form and exhibit difficulties in geometric and boundary treatments. 
Many hybrid methods are proposed to combine the advantages among them.

There are Eulerian and Lagrangian hybrid methods~\cite{foster1996realistic, zhu2005animating, jiang2015affine} as well as vorticity and velocity hybrid methods~\cite{koumoutsakos2008flow, pfaff2012lagrangian}. In general, traditional methods face challenges due to the limited expressivity of their spatial representations and improvements have been proposed over the years. We focus on using continuous representation with more expressivity. Although our optimization-based approach has some limitations compared to established state-of-the-art methods using explicit representations, it offers advantages in detail preservation and 
\cmy{spatial adaptivity}, which we consider a promising direction for exploration.


\paragraph{Emerging Trends in Simulation using Continuous Representations}
Continuous representations with accurate gradient estimations are attracting wide attention in graphics and vision tasks, such as scene reconstruction~\cite{mildenhall2021nerf, feng2024gaussianV1}, video compression~\cite{chen2021nerv}, and physics spatial-temporal approximation~\cite{karniadakis2021physics}. 
There is also a trend to apply INR and Gaussian functions in simulations.
Many works apply them as mass distribution functions and incorporate them with existing numerical solvers (e.g., Material Point Methods) to support physics-based scene editing or animation~\cite{xie2023physgaussian, feng2023pienerf}.
\citet{deng2023neural} leverages INR to store flow maps and generate non-dissipative results using a grid-based fluid solver with long-term advection.
However, fewer works explore how to solve time-dependent \xing{PDEs} using these novel representations as spatial functions.
The most relevant works to us are Implicit Neural Spatial Representations for PDEs (INSR)~\cite{chen2023implicit} and Neural Monte Carlo Fluids (NMC)~\cite{Jain2024NeuralMC}, where the former applies INR as spatial representations for fluid and soft-body simulations and solves temporal \xing{PDEs} via optimization, while the latter takes advantage of the continuity nature of INR by leveraging the walk on stars method for pressure solving and augment the INR to fit boundary conditions.
Our method takes advantage of the locality and flexibility of Gaussian functions and achieves better efficiency and stability in temporal evolution.

\section{Background}

The motion of incompressible fluid is governed by the Navier-Stokes equations with the divergence-free constraint:
\begin{equation}
    \begin{aligned}
    \frac{\partial\boldsymbol u}{\partial t}+(\boldsymbol u\cdot\nabla)\boldsymbol u&=-\frac 1\rho\nabla p+\nu\nabla^2\boldsymbol u+\boldsymbol g,\\
    \nabla\cdot\boldsymbol u&=0,
    \end{aligned}
    \label{eqn:NS}
\end{equation}
where $\boldsymbol u$ is the velocity field, $\rho$ is density, $p$ is pressure, \xing{$\nu$ is kinematic viscosity,} and $\boldsymbol g$ is acceleration due to external force. 
To compute the time evolution of 
$\boldsymbol u$, numerical solvers typically employ an operator splitting scheme, consisting of two main steps: advection and projection.
In the advection step, the equation $\frac{\partial\boldsymbol u}{\partial t}+(\boldsymbol u\cdot\nabla)\boldsymbol u=0$ is solved by transporting the velocity field to its new positions at the next time step, i.e. $\boldsymbol u^*(\boldsymbol x)\gets\boldsymbol u^{n-1}(\boldsymbol\Psi^{n-1}(\boldsymbol x))$, where $\boldsymbol\Psi^{n-1}$ maps a position at frame $n$ to its corresponding position at frame $n-1$. 
In the projection step, the pressure field $p^*(\boldsymbol x)=\frac{p(\boldsymbol x)}{\rho}$ is solved subject to the divergence-free constraint, i.e., $\nabla\cdot(\boldsymbol u^*-\nabla p^*)=0$. The velocity field at the next frame is then updated as $\boldsymbol u^n(\boldsymbol x)\gets\boldsymbol u^*(\boldsymbol x)-\nabla p^*(\boldsymbol x)$. In addition to the two main steps, the external forces are usually applied to the velocity field before advection, while \xing{viscosity} is introduced to the system between advection and projection.

\section{Gaussian Spatial Representation}

\paragraph{Math Definition} 
Motivated by the high expressiveness of 3D Gaussian Splatting \cite{kerbl3Dgaussians} in reconstruction tasks, we propose Gaussian spatial representation (GSR), 
a continuous, differentiable, and memory-efficient
spatial representation based on combined Gaussian functions.
A $d$-dimension ($d\in\{2,3\}$) Gaussian function $G_i:\mathbb R^d\to\mathbb R$ can be formulated as:
\begin{equation}
    G_i(\boldsymbol x)=\exp\left\{-\frac 12(\boldsymbol x-\boldsymbol\mu_i)^\top\boldsymbol\Sigma_i^{-1}(\boldsymbol x-\boldsymbol\mu_i)\right\},\label{eqn:gaussian-function}
\end{equation}
where $\boldsymbol\mu_i$ is the position of the Gaussian particle $i$, and $\boldsymbol\Sigma_i$ is its covariance matrix. Since $\boldsymbol\Sigma_i^{-1}$ is positive definite, we can further decompose it into
\begin{equation}
    \boldsymbol\Sigma_i^{-1}=\boldsymbol R_i\boldsymbol S_i^{-1}\boldsymbol S_i^{-1}\boldsymbol R_i^\top,
\end{equation}
where $\boldsymbol R_i$ is a $d$-dimension rotation matrix, and $\boldsymbol S_i$ is a positive diagonal matrix. The rotation $\boldsymbol R_i$ can be represented as an angle $\theta_i$ in 2D and a quaternion $\boldsymbol r_i$ in 3D. For convenience of implementation, we store the diagonal elements of $\boldsymbol S_i^{-1}$ (denoted as $\boldsymbol s_i^{-1}$) instead of $\boldsymbol S_i$ for particle $i$. A GSR is a vector field $\tilde{\boldsymbol v}:\mathbb R^d\to\mathbb R^m$ defined as the weighted sum of all Gaussian functions:
\begin{equation}
    \tilde{\boldsymbol v}(\boldsymbol x)=\sum_{i=1}^N\boldsymbol v_iG_i(\boldsymbol x),
    \label{eqn:GSR-pre-def}
\end{equation}
where $\boldsymbol v_i\in\mathbb R^m$ is the weight of Gaussian particle $i$, and $N$ is the number of Gaussian particles. Hence, the parameters of a GSR are $\Theta=\{\boldsymbol v_i,\boldsymbol\mu_i,\theta_i,\boldsymbol s_i^{-1}:i=1,\cdots,N\}$ in 2D and $\Theta=\{\boldsymbol v_i,\boldsymbol\mu_i,\boldsymbol r_i,\boldsymbol s_i^{-1}:i=1,\cdots,N\}$ in 3D.

\paragraph{Efficiency Improvements} 
The GSR defined in Equation~\ref{eqn:GSR-pre-def} initially requires $O(N)$ floating-point operations to evaluate the field at a single point, which becomes computationally prohibitive as the number of queries grows. 
To mitigate this, we apply a local restriction to each Gaussian function, exploiting its rapid decay with distance. This results in the clamped Gaussian function formulation:
\begin{equation}
    \hat G_i(\boldsymbol x)=
    \begin{cases}
        G_i(\boldsymbol x)-c,&G_i(\boldsymbol x)\ge c\\
        0,&G_i(\boldsymbol x)<c
    \end{cases},
\end{equation}
where $c$ is a small positive threshold. We subtract the Gaussian function by $c$ to avoid discontinuity of the kernel function. Accordingly, we change the definition of GSR in Equation~\ref{eqn:GSR-pre-def} into:
\begin{equation}
    \tilde{\boldsymbol v}(\boldsymbol x)=\sum_{i=1}^N\boldsymbol v_i\hat G_i(\boldsymbol x).
    \label{eqn:GSR-def}
\end{equation}
A hash table is employed to store the Gaussian particles based on spatial locality, enabling fast retrieval of only the relevant Gaussian particles. This reduces the time complexity for each query to $O(1)$. Details can be found in supplemental files.

\paragraph{Advantage on Differentiability} 
The gradient of the GSR can be computed directly from its definition:
\begin{equation}
    \nabla\tilde{\boldsymbol v}(\boldsymbol x)=\sum_{i=1}^N\boldsymbol v_i\nabla\hat G_i(\boldsymbol x)=-\sum_{i\in\mathcal N(\boldsymbol x)}G_i(\boldsymbol x)\boldsymbol v_i(\boldsymbol x-\boldsymbol\mu_i)^\top\boldsymbol\Sigma_i^{-1},
    \label{eqn:GSR-grad}
\end{equation}
\xing{where $\mathcal N(\boldsymbol x)=\{i:\hat G_i(\boldsymbol x)\ge 0\}$ is the set of the particle indices near $\boldsymbol x$.}
Though the GSR is not differentiable at the boundary of any clamped Gaussian function, i.e. $\tilde{\boldsymbol v}(\boldsymbol x)$ is not differentiable when $\exists i\in\{1,\cdots,N\}$, $G_i(\boldsymbol x)=c$, Equation~\ref{eqn:GSR-grad} is a good approximation for small $c$s.
Furthermore, we can naturally derive the divergence and curl operators from this formulation.
Unlike implicit neural representations that rely on auto-differentiation for computing differential quantities, GSR allows direct and efficient differentiation, with the same time complexity as evaluating the GSR field itself. This inherent advantage leads to faster optimization when enforcing physics-based constraints, making GSR a computationally superior choice for dynamic simulations.
%

\section{Algorithm}

Our work focuses on the \xing{dynamics} of inviscid fluids with no external force, i.e. $\nu=0$ and $\boldsymbol g=\boldsymbol 0$ in Equation~\ref{eqn:NS}. Our fluid solver can be divided into initialization, \xing{physics-based optimization}, and reseeding, as shown in Algorithm~\ref{alg:fluid-solver}.

\begin{algorithm}
\caption{Fluid solver with GSR}
\label{alg:fluid-solver}
\begin{algorithmic}[1]
\STATE $u^0 \leftarrow \text{Initialize}(u)$
\FOR{$n \leftarrow 1, \cdots, T$}
\STATE $\tilde{u}^{n-1} \leftarrow \text{Reseed}(u^{n-1})$
\newline //An initial guess for physics-based optimization:
\STATE $\tilde{u}^* \leftarrow \text{AdvectPositions}(\tilde{u}^{n-1})$
\newline //Physics-based optimization:
\STATE $u^n \leftarrow \text{OptimizeLosses}(\tilde{u}^*, u^{n-1})$
\ENDFOR
\end{algorithmic}
\end{algorithm}

\subsection{Initialization}
\label{sec:fit-GSR}


At the very beginning of the simulation, we initialize the GSR $\tilde{\boldsymbol u}^0(\boldsymbol x)$ to fit a given velocity field $\boldsymbol u(\boldsymbol x)$.
GSR is capable of fitting any continuous vector field if trained properly with a sufficient number of kernels.
For any vector field $\boldsymbol v:\mathbb R^d\to\mathbb R^m$ defined on domain $\mathcal D\subset\mathbb R^d$, our goal is to find a GSR $\tilde{\boldsymbol v}:\mathbb R^d\to\mathbb R^m$ as close to $\boldsymbol v(\boldsymbol x)$ on $\mathcal D$ as possible. This can be interpreted as an optimization problem: $\argmin_\Theta\frac 1{d\vert\mathcal D\vert}\int_\mathcal D\|\boldsymbol v(\boldsymbol x)-\tilde{\boldsymbol v}(\boldsymbol x)\|_1\mathrm dV$.
Directly evaluating the \xing{integral} is difficult, thus we take the following value loss to approximate it in a Monte-Carlo way:
\begin{equation}
    \mathcal L_\mathrm{val}=\frac 1{Qd}\sum_{j=1}^Q\|\boldsymbol v(\boldsymbol x_j)-\tilde{\boldsymbol v}(\boldsymbol x_j)\|_1,
\end{equation}
where $\boldsymbol x_1,\cdots,\boldsymbol x_Q$ are \xing{uniformly} randomly sampled from $\mathcal D$ \xing{in each iteration}.
To promote high-quality fitting of GSR, we employ an additional gradient loss:
\begin{equation}
    \mathcal L_\mathrm{grad}=\frac 1{Qd^2}\sum_{j=1}^Q\|\nabla\boldsymbol v(\boldsymbol x_j)-\nabla\tilde{\boldsymbol v}(\boldsymbol x_j)\|_\mathrm{sum},
\end{equation}
where $\|\cdot\|_\mathrm{sum}=\sum_{k=1}^d\sum_{l=1}^m|[\cdot]_{kl}|$ is the sum of the absolute \cmy{values} of all elements. 
The gradient loss can supervise a local neighborhood, which makes a difference to the point-wise value loss, improving the training efficiency on the continuous field.

In addition to the value loss and the gradient loss,
we leverage anisotropic loss $\mathcal L_\mathrm{aniso}$ proposed by \citet{xie2023physgaussian} and volume loss $\mathcal L_\mathrm{vol}$ proposed by \cmy{\citet{feng2024gaussianV1}} as regularization terms:
\begin{align}
    &\mathcal L_\mathrm{aniso}=\frac 1N\sum_{i=1}^N\max\left(\frac{\max(\boldsymbol s_i)}{\min(\boldsymbol s_i)}-r_\mathrm{aniso}, 0\right)~,\text{and}\\
    &\mathcal L_\mathrm{vol}=\frac 1N\sum_{i=1}^N\left(\frac{\prod(\boldsymbol s_i)}{\frac 1N\sum_{j=1}^N\prod(\boldsymbol s_j)}-1\right)^2,
\end{align}
where $\max(\cdot)$ is the maximum element of a vector, $\prod(\cdot)$ is the product of all elements of a vector, $r_\mathrm{aniso}$ is the maximum threshold of the ratio between the major axis length and minor axis length of a Gaussian particle. We take $r_\mathrm{aniso}=1.5$ in all our experiments.

The total loss in the initialization process is: 
\begin{equation}
    \mathcal L_\mathrm{init}=\mathcal L_\mathrm{val}+\mathcal L_\mathrm{grad}+\mathcal L_\mathrm{aniso}+\mathcal L_\mathrm{vol}.
    \label{eqn:init-loss}
\end{equation}
We initialize the parameters of GSR adaptively according to the particle number, with details specified in the supplementary. We then run the optimization for enough iterations to get an initial GSR ready for the simulation.

\subsection{\xing{Physics-Based Optimization}}

\xing{We formulate the time integration as an optimization problem with physics-based losses, training the GSR from an initial guess inspired by Lagrangian advection to }\cmy{better}
\xing{represents the velocity field at the next frame.}
Our optimization process leverages physics-guided gradients to ensure adherence to physical constraints. 
By employing a sampling-based strategy, our method effectively handles boundary conditions without requiring explicit cell-cutting operations, offering a seamless and flexible approach to simulate complex geometries. 
\xing{The details of the initial guess, losses, gradient adjustments, and boundary sampling are described as follows.}

\subsubsection{\xing{Advection-Based Initial Guess}}
\xing{To encourage fast convergence and temporal consistency, we make an initial guess for the physics-based optimization with an approximated velocity field advected from the last frame.}
\xing{Drawing inspiration from the advection step in Lagrangian methods, we treat the Gaussian particles as Lagrangian particles and update} the position of each particle by moving it along the velocity for a time step $h$. 
\cmy{Unlike Lagrangian approaches where each particle moves with its own velocity, our Gaussian particles are advected using velocities sampled from the Gaussian field according to Eq.~\ref{eqn:GSR-pre-def}.}
To improve accuracy, we apply the 4-th order Runge-Kutta convention. Denoting $\boldsymbol\Phi^{n-1}:\mathbb R^d\to\mathbb R^d$ as the mapping from a position at time step $n-1$ to the position after applying the RK4 time integration, the advection step is given by $\boldsymbol\mu_i^*\gets\boldsymbol\Phi^{n-1}(\boldsymbol\mu_i^{n-1})$.

While this \xing{step} does not precisely solve $\frac{\partial\boldsymbol u}{\partial t}+(\boldsymbol u\cdot\nabla)\boldsymbol u=0$, it provides a reliable initial state for subsequent steps. It maintains a relatively uniform particle distribution without clustering, as long as the velocity field remains approximately divergence-free, as ensured in later steps. 
We also use the advected positions as regularization in subsequent optimization by introducing a constraint specified in Section~\ref{sec:position-penalty} to limit excessive drift of particles. Serving both as initialization and regularization, the advected positions accelerate optimization by guiding the solution toward a local optimum, improving temporal consistency and stability of the spatial representation. The effectiveness of \xing{this} step is validated through an ablation study in Section~\ref{sec:Ablation}.

\subsubsection{The PDE Losses}

We first introduce the vorticity loss and the divergence loss \cmy{used in the physics-based optimization step}:
\begin{align}
    &\mathcal L_\mathrm{vor}=\frac 1{Q\hat d}\sum_{j=1}^Q\|\nabla\times\tilde{\boldsymbol u}^n(\boldsymbol x_j)-\omega(\boldsymbol x_j)\|_1,\\
    &\mathcal L_\mathrm{div}=\frac 1Q\sum_{j=1}^Q|\nabla\cdot\tilde{\boldsymbol u}^n(\boldsymbol x_j)|^2,
\end{align}
where $\hat d=1$ in 2D examples and \xing{$\hat d=3$} in 3D, \xing{$\boldsymbol x_1,\cdots,\boldsymbol x_Q$ are uniformly randomly sampled from $\mathcal D$ in each iteration of the optimization,} $\omega(\boldsymbol x)$ is the vorticity field advected from $\nabla\times\tilde{\boldsymbol u}^{n-1}(\boldsymbol x)$. In 2D, the vorticity simply transports along the velocity field, i.e.
\begin{equation}
    \omega(\boldsymbol x)=\nabla\times\tilde{\boldsymbol u}^{n-1}(\boldsymbol\Psi^{n-1}(\boldsymbol x)).
\end{equation}
In 3D, the vorticity field evolves according to $\frac{\mathrm D\omega}{\mathrm Dt}=\nabla\boldsymbol u\cdot\omega$, which can be characterized by the bidirectional flow map as mentioned by \citet{wang2024eulerian}:
\begin{equation}\small
    \omega(\boldsymbol x)=\mathrm d\boldsymbol\Phi^{n-1}(\boldsymbol\Psi^{n-1}(\boldsymbol x))\nabla\times\tilde{\boldsymbol u}^{n-1}(\boldsymbol\Psi^{n-1}(\boldsymbol x)),
    \label{eqn:advect-vorticity-explicit}
\end{equation}
where $\mathrm d\boldsymbol\Phi^{n-1}(\boldsymbol x)$ denotes the Jacobian matrix of the forward mapping $\boldsymbol\Phi^{n-1}$ at $\boldsymbol x$.

\subsubsection{Gradient Projection}
\label{sec:grad-proj}

\begin{figure}
    \centering
    \includegraphics[width=0.25\textwidth]{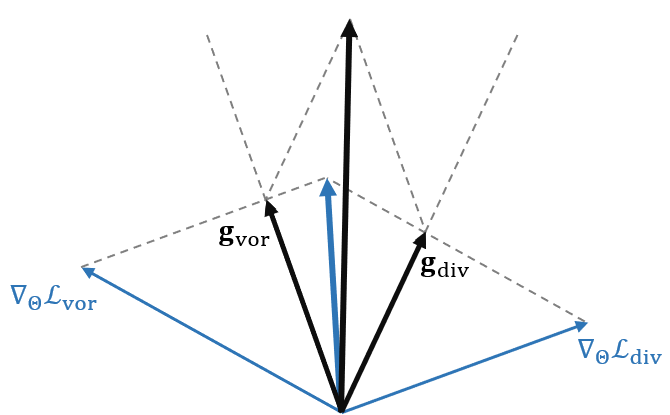}
    \caption{The gradiant projection technique can turn the contradicted gradients into compatible ones, resulting in larger steps in gradient descent.}
    \label{fig:grad-proj}
\end{figure}

\begin{figure*}[htbp]\small
\centering
\newcommand{\formattedgraphics}[1]{\includegraphics[trim=110 70 170 85,clip,width=\linewidth]{#1}}
\newcommand{\negspace}{\hspace{-1pt}}
\newcommand{\negvspace}{\vspace{-18pt}}
\begin{subfigure}{0.16\textwidth}
        \centering
        \formattedgraphics{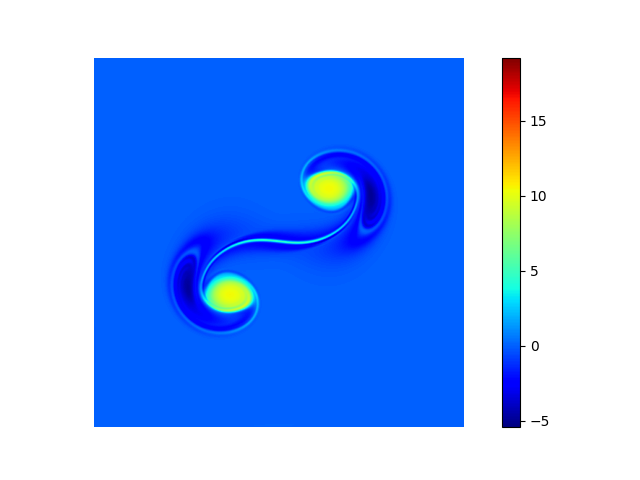}\negvspace
        \caption{Euler}
        \label{fig:taylor-euler}
\end{subfigure}
 \negspace
 \begin{subfigure}{0.16\textwidth}
        \centering
        \formattedgraphics{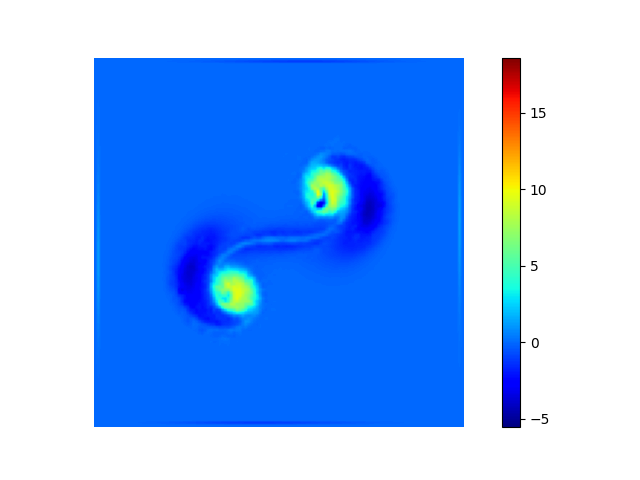}\negvspace
        \caption{FLIP}
        \label{fig:taylor-flip}
\end{subfigure}
 \negspace
 \begin{subfigure}{0.16\textwidth}
        \centering
        \formattedgraphics{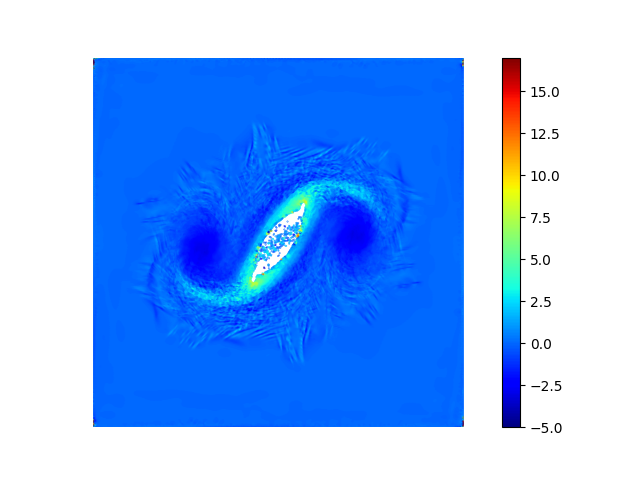}\negvspace
        \caption{SPH}
        \label{fig:taylor-sph}
\end{subfigure}
 \negspace
 \begin{subfigure}{0.16\textwidth}
        \centering
        \formattedgraphics{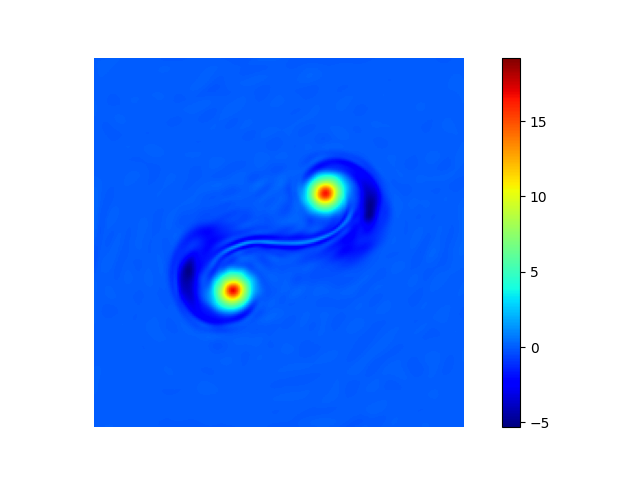}\negvspace
        \caption{INSR}
        \label{fig:taylor-INSR}
\end{subfigure}
 \negspace
 \begin{subfigure}{0.16\textwidth}
        \centering
        \formattedgraphics{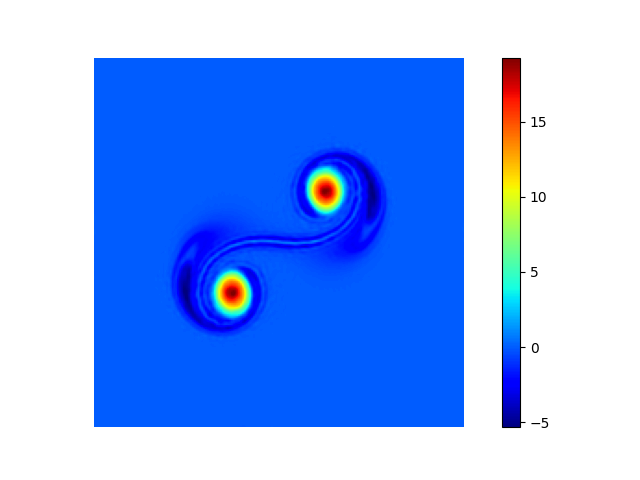}\negvspace
        \caption{Ours}
        \label{fig:taylor-ours}
\end{subfigure}
 \negspace
 \begin{subfigure}{0.16\textwidth}
        \centering
        \formattedgraphics{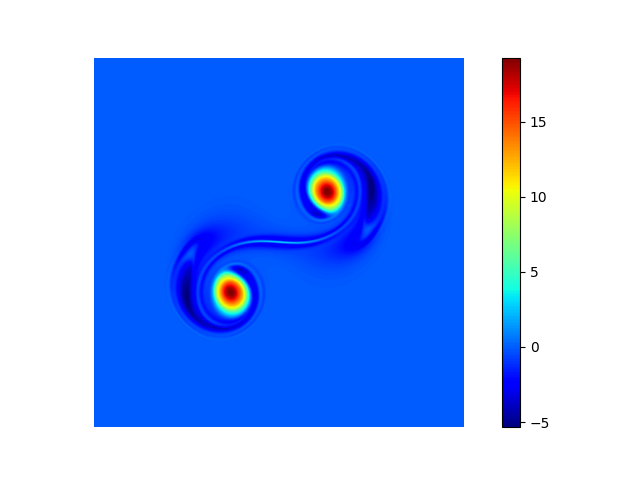}\negvspace
        \caption{Ground-Truth}
        \label{fig:taylor-GT}
\end{subfigure}\vspace{-6pt}
\caption{\small Our method compared with \xing{traditional Eulerian and Lagrangian} discretizations and INSR on the Taylor vortex example. We show frame 300 while more are given in figure~\ref{fig:comparison}.}
\label{fig:taylor-cmp}
\end{figure*}

We apply the gradient projection technique, a gradient strategy typically applied in \textit{Multi-Task Learning (MTL)}~\cite{Yu2020, Dong2022,liu2025config}, to the vorticity loss and the divergence loss in the pressure solve training process, as their gradients may contradict. In the backward stage, after computing the gradients of the two losses $\nabla_\Theta\mathcal L_\mathrm{vor}$ and $\nabla_\Theta\mathcal L_\mathrm{div}$, we check if their dot product is negative. If so, this indicates a contradiction between the two gradients, meaning that following one gradient would increase the other loss. When this occurs, let $\boldsymbol t_1,\boldsymbol t_2$ be the normalized vector of $\nabla_\Theta\mathcal L_\mathrm{vor}$ and $\nabla_\Theta\mathcal L_\mathrm{div}$, respectively. We then modify the gradients of the two losses to:
\begin{align}
    &\boldsymbol g_\mathrm{vor}=\nabla_\Theta\mathcal L_\mathrm{vor}-(\nabla_\Theta\mathcal L_\mathrm{vor}\cdot\boldsymbol t_2)\boldsymbol t_2,\\
    &\boldsymbol g_\mathrm{div}=\nabla_\Theta\mathcal L_\mathrm{div}-(\nabla_\Theta\mathcal L_\mathrm{div}\cdot\boldsymbol t_1)\boldsymbol t_1.
\end{align}
Moving along $\boldsymbol g_\mathrm {vor}$ would not affect $\mathcal L_\mathrm{div}$ and vice versa, hence increasing the efficiency of each gradient descent step, as shown in Figure~\ref{fig:grad-proj}. This technique follows the MTL strategy proposed by \citet{Yu2020}, which in our case can reduce the ripple artifact in the vorticity field due to sub-optimal optimization processes converging to local minima, as later shown by an ablation test in Section~\ref{sec:Ablation}. 

\subsubsection{Boundary Handling}
There are two types of boundary conditions across all experiments in this paper. The First type can be formulated as $\boldsymbol u(\boldsymbol x)=\boldsymbol u_\mathrm b(\boldsymbol x)$ while the second $(\boldsymbol u\cdot\boldsymbol n)(\boldsymbol x)=f(\boldsymbol x)$, where $\boldsymbol x\in\partial\mathcal D$, $\boldsymbol n$ is the boundary normal, $\boldsymbol u_\mathrm b$ and $f$ are given functions.
They stand for the "no-slip" and "free-slip" conditions, respectively.
We handle the two types of boundary conditions by introducing two boundary losses:
\begin{align}
    &\mathcal L_\mathrm{b1}=\frac 1{Q_\mathrm{b1}d}\sum_{j=1}^{Q_\mathrm{b1}}\|\tilde{\boldsymbol u}^n(\boldsymbol y_j)-\boldsymbol u_\mathrm b(\boldsymbol y_j)\|_1,\\
    &\mathcal L_\mathrm{b2}=\frac 1{Q_\mathrm{b2}}\sum_{j=1}^{Q_\mathrm{b2}}\vert\tilde{\boldsymbol u}^n(\boldsymbol z_j)\cdot\boldsymbol n_j-f(\boldsymbol z_j)\vert,
\end{align}
where $\boldsymbol y_1,\cdots\boldsymbol y_{Q_\mathrm{b1}},\boldsymbol z_1,\cdots,\boldsymbol z_{Q_\mathrm{b2}}$ are uniformly \xing{randomly} sampled from the corresponding type of the domain boundary \xing{in each iteration of the optimization}, $\boldsymbol n_j=\boldsymbol n(\boldsymbol z_j)$ are the normal vectors at the sampled points.

\subsubsection{The Position Penalty}\label{sec:position-penalty}
To sufficiently exploit the initial distribution provided by the previous advection step and prevent the particles from clustering, we add another regularization term to \xing{constrain} their positions:
\begin{equation}
    \mathcal L_\mathrm{pos}=\frac 1{Nd}\sum_{i=1}^N\|\boldsymbol\mu^n_i-\boldsymbol\mu^*_i\|^2.
\end{equation}

The total loss \cmy{optimizing the temporal evolution }
is the weighted combination of the vorticity loss, the divergence loss, the boundary loss, and the regularization terms:
{\small
\begin{equation}
    \mathcal L =\mathcal L_\mathrm{vor}+\lambda_\mathrm{div}\mathcal L_\mathrm{div}+\lambda_\mathrm{b1}\mathcal L_\mathrm{b1}+\lambda_\mathrm{b2}\mathcal L_\mathrm{b2}+\lambda_\mathrm{aniso}\mathcal L_\mathrm{aniso}+\lambda_\mathrm{vol}\mathcal L_\mathrm{vol}+\lambda_\mathrm{pos}\mathcal L_\mathrm{pos}.  \nonumber
\end{equation}}
We then use the Adam to optimize the total loss, with the gradient of $\mathcal L_\mathrm{vor}$ and $\mathcal L_\mathrm{div}$ replaced by $\boldsymbol g_\mathrm{vor}$ and $\boldsymbol g_\mathrm{div}$ respectively.

\subsection{Reseeding}
\label{sec:reseeding}


Although the anisotropic regularization term is applied during projection, some Gaussian particles may inevitably become excessively elongated due to turbulent fluid motion. 
To address this, we introduce a reseeding procedure at the beginning of each time step. We split particles whose maximum scale is at least twice its minimum scale, i.e. split particle $i$ if \xing{$\max(\boldsymbol s_i)\ge r_\mathrm{aniso}\min(\boldsymbol s_i)$}. The positions of the two new particles resulting from the split are sampled from the Gaussian distribution $\mathcal N(\boldsymbol\mu_i^{n-1},\boldsymbol\Sigma_i^{n-1})$. Their maximum scales are halved, while the other parameters of the new particles are inherited from particle $i$. This is followed by a local fitting procedure where we optimize only the parameters of the new particles and their neighbors using the same loss function as during initialization (i.e. $\mathcal L_\mathrm{init}$ from Equation~\ref{eqn:init-loss}).
\xing{Note the number of split particles is influenced by $\lambda_\mathrm{aniso}$, as a larger $\lambda_\mathrm{aniso}$ imposes a stronger penalty over the particles' anisotropy in the previous time step, reducing the number of elongated particles.}

\section{Results}

{
\setlength{\tabcolsep}{0.1pt} 
\begin{table}[htbp]\footnotesize
\caption{\small Performance comparison. \xing{Timestep is measured in seconds.} Running time (in seconds) indicates the solvers' average time cost. Memory usage indicates the average size (in KB) taken by the spatial representation.}\vspace{-8pt}
\begin{tabular}{c|c|c|c|c|c}
\hline
\textbf{Example}                      & \textbf{Timestep}      & \textbf{Method}    & \textbf{Running Time} & \textbf{Particle No.} & \textbf{Mem. Usage} \\[2pt] \hline
\multirow{3}{*}{\xing{Taylor-Green}}  & \multirow{3}{*}{\xing{0.001}} & \xing{INSR}        & \xing{$403$}       & \xing{-}                & \xing{$32.1$}          \\ \cline{3-6}
                                      &                        & \xing{NMC}         & \xing{$39$}        & \xing{-}                & \xing{$103.8$}         \\ \cline{3-6}
                                      &                        & \xing{Ours}        & \xing{$38$}        & \xing{$576$}            & \xing{$17.9$}          \\ \hline
\multirow{2}{*}{Taylor vortex}        & \multirow{2}{*}{0.01}  & INSR               & $378$              & -                       & $32.1$                 \\ \cline{3-6}
                                      &                        & Ours               & $63$               & $5041\sim 5511$         & $148.5$                \\ \hline
Leapfrog 2D                           & 0.025                  & Ours               & $48$               & $4846\sim 5041$         & $137.0$                \\ \hline
Vortices pass                         & \xing{0.01}            & Ours               & \xing{$47$}        & \xing{$5041\sim 5273$}  & \xing{$143.1$}         \\ \hline
\multirow{2}{*}{Karman vortex street} & \multirow{2}{*}{0.05}  & NMC                & $81$               & -                       & $134.3$                \\ \cline{3-6}
                                      &                        & Ours               & $214$              & $20408\sim 24001$       & $598.6$                \\ \hline
Leapfrog 3D                           & 0.02                   & Ours               & $228$              & $64000$\textsuperscript{\dag}     & $3252.2$               \\ \hline
Ring collide                          & 0.02                   & Ours               & $206$              & $64000$\textsuperscript{\dag}     & $3252.2$               \\ \hline
Smoking bunny                         & 0.02                   & Ours               & $201$              & $64000$\textsuperscript{\dag}     & $3252.2$               \\ \hline
\end{tabular}
\footnotesize\textsuperscript{\dag} In all 3D examples, particle splitting does not occur due to sufficient initial particles.
\label{tab:performance}
\end{table}
}

\begin{figure*}[htbp]
    \centering
    \newcommand{\formattedgraphics}[1]{\includegraphics[trim=100 50 230 50,clip,width=\columnwidth]{#1}}
    \begin{subfigure}{0.31\textwidth}
        \centering
        \formattedgraphics{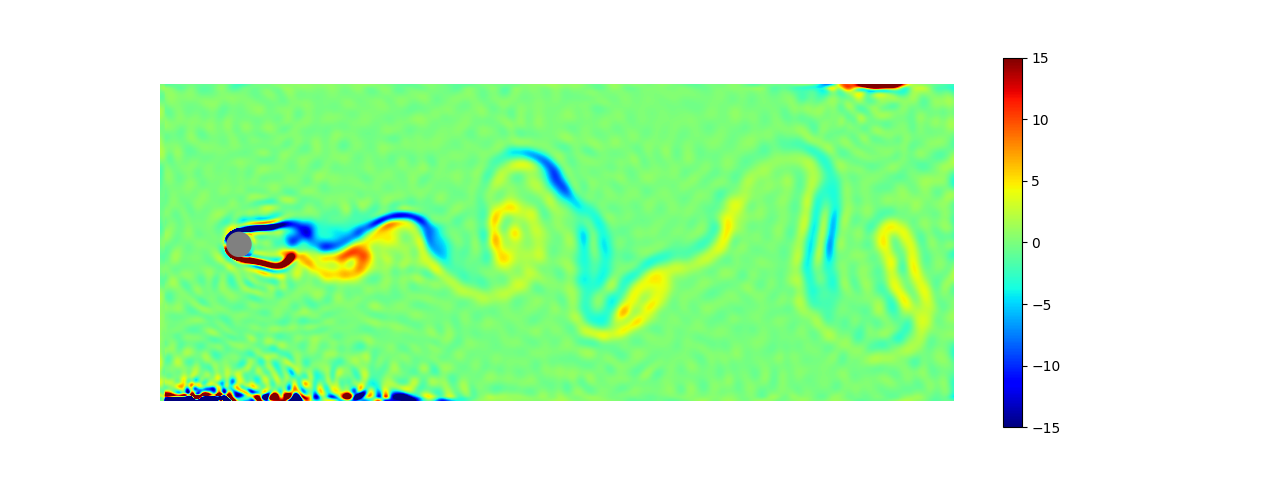}\vspace{-18pt}
        \caption{NMC}
        \label{fig:karman-NMC}
    \end{subfigure} ~
    \begin{subfigure}{0.31\textwidth}
        \centering
        \formattedgraphics{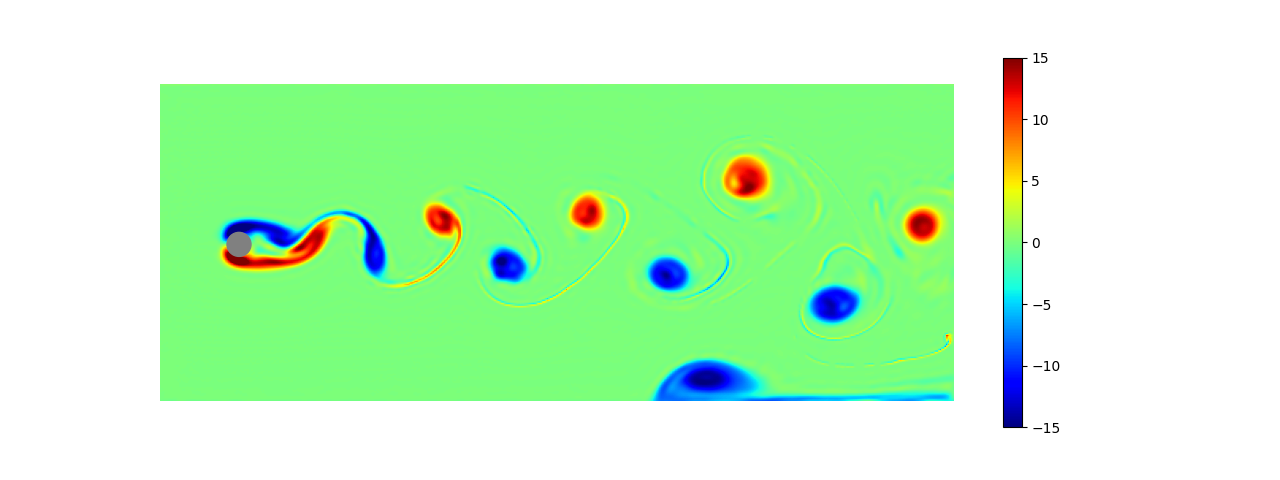}\vspace{-18pt}
        \caption{Ours}
        \label{fig:karman-ours}
    \end{subfigure} ~
    \begin{subfigure}{0.31\textwidth}
        \centering
        \formattedgraphics{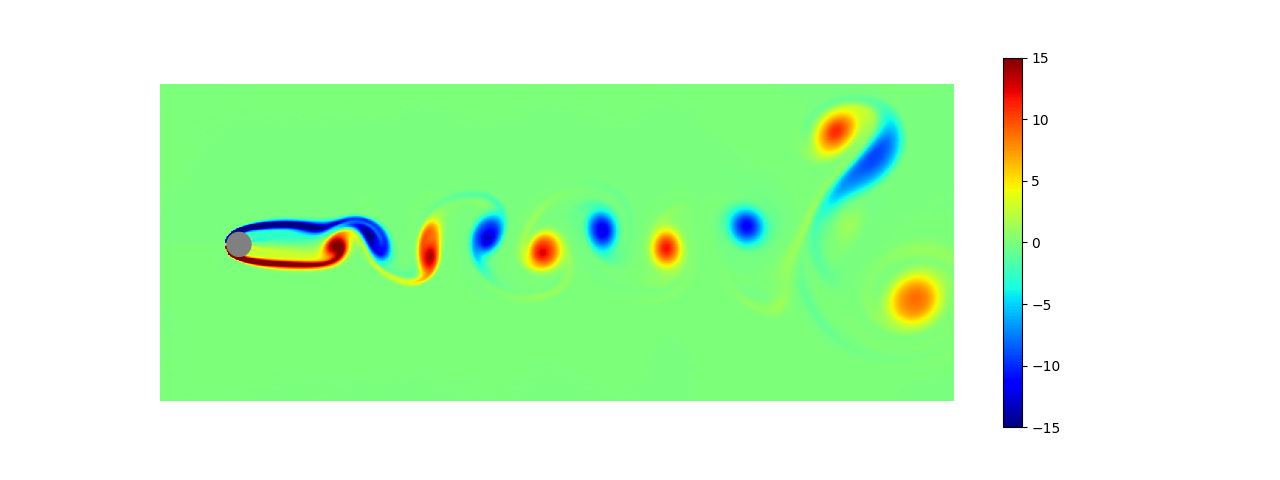}\vspace{-18pt}
        \caption{SF+R (Reference)}
        \label{fig:karman-SF+R}
    \end{subfigure}
    \vspace{-6pt}
    \caption{\footnotesize Karman vortex street example by Neural Monte Carlo (NMC), our method, and stable fluids with reflection projection (SF+R) proposed by \citet{Zehnder2018Reflection}. The sub-figures display the vorticity fields of frame 152, 199 and 199 of the simulation results, respectively.}\vspace{-8pt}
    \label{fig:karman}
\end{figure*}
\begin{figure}[htbp]
      \centering
    \includegraphics[trim=50 84 50 100,clip,width=\columnwidth]{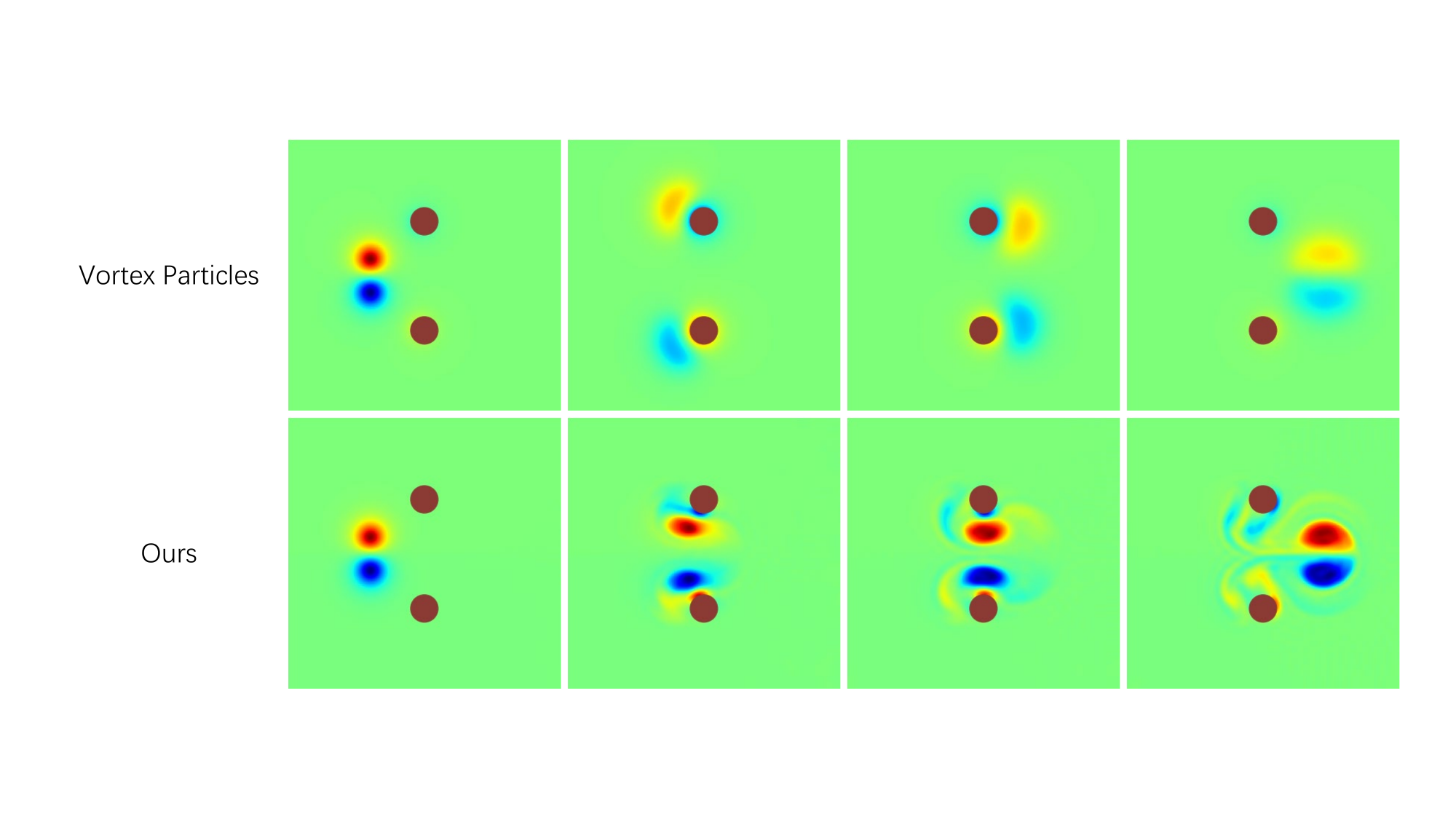}
    \vspace{-24pt}
    \caption{\footnotesize Simulations of the vortices pass example produced by \xing{vortex particles with conserved harmonic components} and our method. A pair of vortices pass through a gap between two spherical smooth objects. The images from left to right are showing frames 0, \xing{150}, \xing{270} and \xing{400}, respectively.}
    \label{fig:vortices_pass}
\end{figure}

We evaluate our method on a diverse set of 2D and 3D examples, with all results provided in the supplemental video. Notably, unlike most methods based on first-order optimization, our approach does not require hyperparameter tuning across different scenarios, highlighting the robustness of GSR. To handle examples of varying scales, we rescale the entire fluid domain, boundary geometries, and initial velocity to a canonical size. With this adjustment, a single parameter setting suffices for all 2D examples and another for all 3D examples. Further details on hyperparameter settings, the normalization strategy, and scene configurations are available in the supplemental document.

\xing{We first show the numerical accuracy and convergence rate of GSR with a quantitative study on an example with analytical solution.}
We \xing{then} validate its effectiveness by demonstrating the adaptive spatial accuracy on classical fluid phenomena. Next, we evaluate the stability of our optimization-based solver, in together with its boundary handling. \xing{Moreover}, we demonstrate that GSR effectively preserves vortices and is capable of handling complex dynamic behaviors with highly intricate fluid simulations. Finally, multiple ablations are conducted to assess the contribution of key components in our method. The performance for all examples is provided.

\subsection{\xing{Quantitative Study}}

\xing{We conduct a quantitative study using the Taylor-Green vortex experiment, analyzing the numerical accuracy of our method and the convergence rates of the optimizations during initialization and time integration. In this example, the initial velocity field is set to
\begin{equation}
    \boldsymbol u(x,y)=\begin{bmatrix}\sin x\cos y\\-\cos x\sin y\end{bmatrix}
\end{equation}
defined on the fluid domain $\mathcal D=[0,2\pi]\times[0,2\pi]$. We then run the simulation for $100$ frames with a $0.001$ seconds timestep.}

\begin{figure}[htbp]\footnotesize
    \centering
    \begin{minipage}{\columnwidth}
        \centering
        \newcommand{\formattedgraphics}[1]{\includegraphics[trim=60 30 50 20,clip,width=0.27\textwidth]{#1}}
        \begin{tabular}{@{}c@{}c@{}c@{}c@{}}
            \includegraphics[trim=110 40 100 50,clip,width=0.205\textwidth]{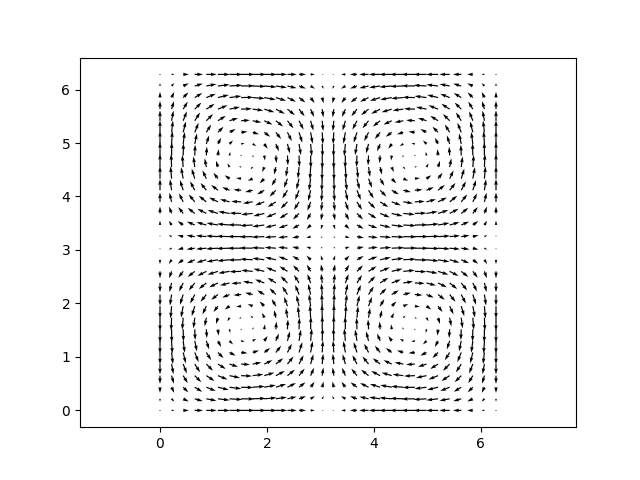} &
            \formattedgraphics{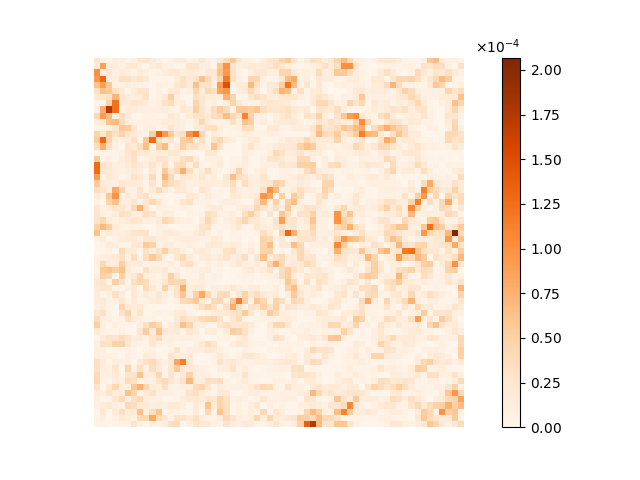} &
            \formattedgraphics{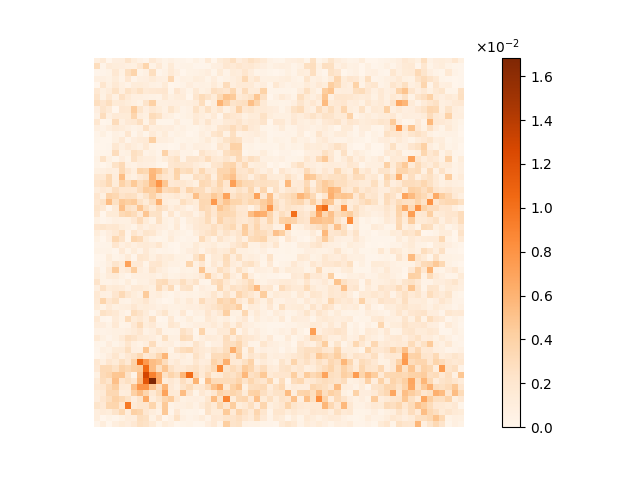} &
            \formattedgraphics{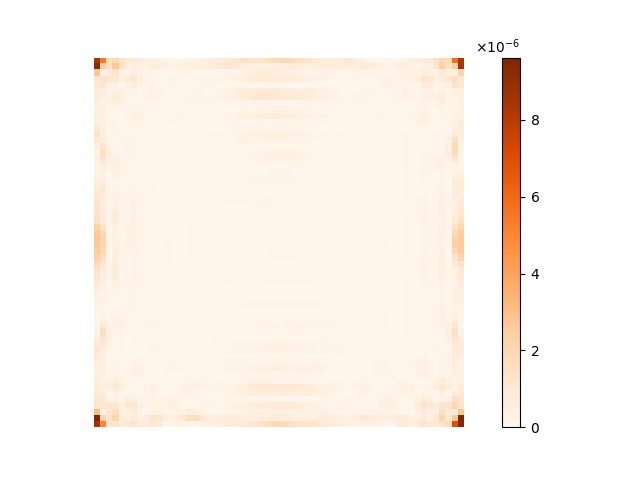} \\
            GT Vel.& INSR & NMC & Ours
        \end{tabular}
    \end{minipage}
    \begin{minipage}{\columnwidth}
        \centering
        \xing{
        \begin{tabular}{c|cccc}
            \textbf{Frame} & \textbf{Eulerian} & \textbf{INSR} & \textbf{NMC} & \textbf{Ours} \\ \hline
            $0$ & $2.432\times 10^{-4}$ & $8.998\times 10^{-7}$ & $1.829\times 10^{-4}$ & $9.957\times 10^{-8}$ \\ 
            $50$ & $9.757\times 10^{-3}$ & $1.715\times 10^{-5}$ & $6.492\times 10^{-4}$ & $2.510\times 10^{-7}$ \\ 
            $100$ & $2.019\times 10^{-2}$ & $1.992\times 10^{-5}$ & $1.725\times 10^{-3}$ & $2.181\times 10^{-7}$
        \end{tabular}
        }
    \end{minipage}
    \caption{\xing{Quantitative comparison using the Taylor-Green vortex example. Top-left: velocity field of the example. Top-right: the MSE fields at frame $100$ of different methods. Bottom: MSE between the simulated velocity fields of different methods and the ground truth.}}
    \label{fig:taylor_green-quantitative}
\end{figure}

\xing{Since the velocity field will remain constant on inviscid incompressible fluids, we measure the numerical error of different solvers by the mean squared error (MSE) between the simulated velocity fields and $\boldsymbol u(x,y)$ sampled on a $60\times 60$ uniform grid at certain frames. The results in Figure~\ref{fig:taylor_green-quantitative} show that our method has significantly lower numerical error than INR-based methods, even with less memory usage as in Table~\ref{tab:performance}. We also conduct a comparison against the semi-implicit Eulerian method run on a $64\times 64$ MAC-grid (taking up $49$KB each frame), which has the highest error among all as in Figure~\ref{fig:taylor_green-quantitative}.}

\begin{figure}
    \centering
    \begin{subfigure}{0.5\columnwidth}
        \centering
        \includegraphics[width=\textwidth]{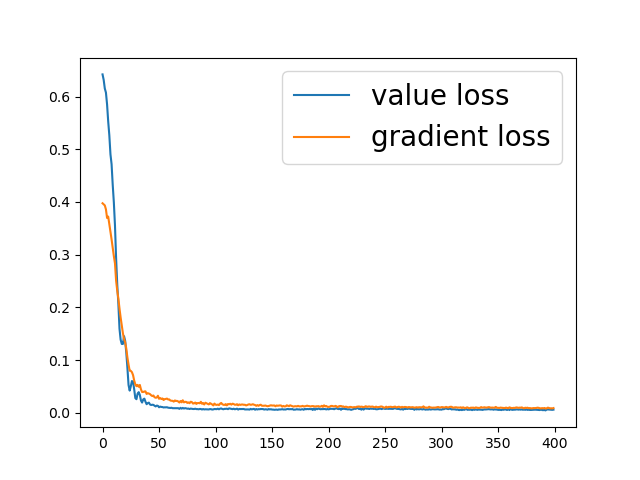}
    \end{subfigure}\hfill
    \begin{subfigure}{0.5\columnwidth}
        \centering
        \includegraphics[width=\textwidth]{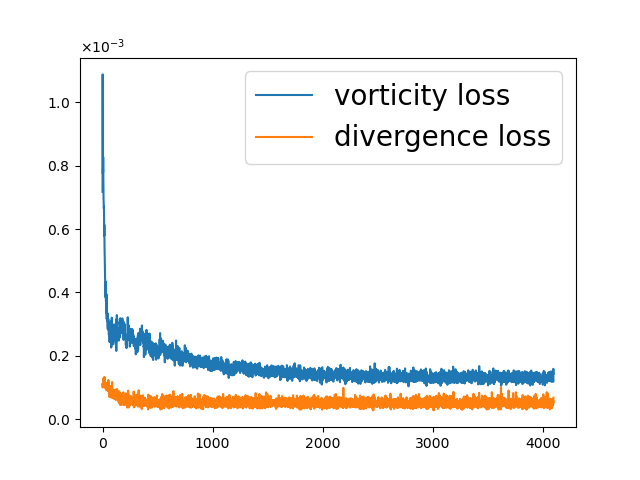}
    \end{subfigure}
    \caption{\xing{Loss curves in the optimizations during initialization (left) and time integration (right).}}
    \label{fig:taylor_green-convergence}
\end{figure}

\xing{We plot the loss curves of the optimizations during initialization and a time integration process with time step size $0.01$ seconds from the $100$-th frame in Figure~\ref{fig:taylor_green-quantitative}, which indicates fast convergence of our method. \cmy{Figure~\ref{fig:taylor_green-convergence}} shows the loss curves for $400$ iterations of the initialization optimization \cmy{and $4100$ iterations of the time-integration optimization}. 
During simulation, we apply an early stop on the time-integration optimization if both vorticity and divergence loss do not change significantly for $500$ iterations.}

\subsection{Validation}

\begin{figure*}[htbp]\small
\centering
\newcommand{\formattedgraphics}[1]{\includegraphics[trim=110 50 170 65,clip,width=0.175\textwidth]{#1}}
\newcommand{\formattedgraphicslst}[1]{\includegraphics[trim=100 50 144 65,clip,width=0.21\textwidth]{#1}}
\newcommand{\negspace}{\hspace{-1.8pt}}
\newcommand{\negvspace}{\vspace{-6pt}}
\begin{tabular}{p{2.5cm} l}  
  \cmy{\textbf{Semi-Implicit Euler\newline $512\times 512$ \newline ($2$ MB)}} & 
  \begin{minipage}{0.8\textwidth}\negvspace
    \formattedgraphics{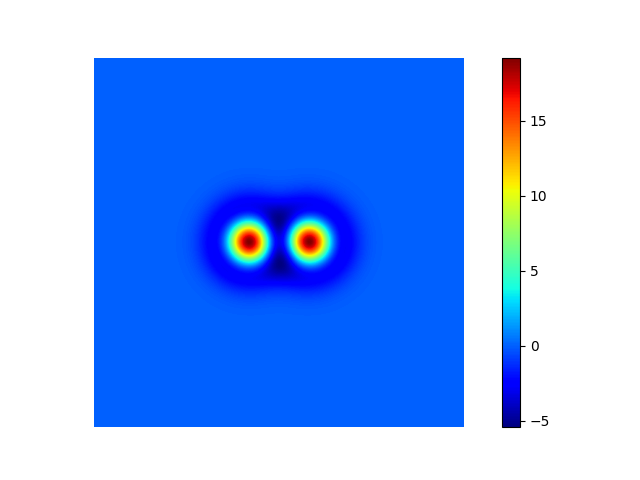}
    \negspace
    \formattedgraphics{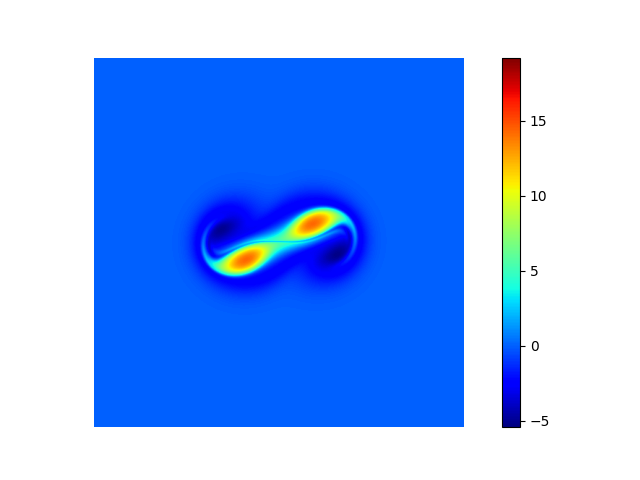}
    \negspace
    \formattedgraphics{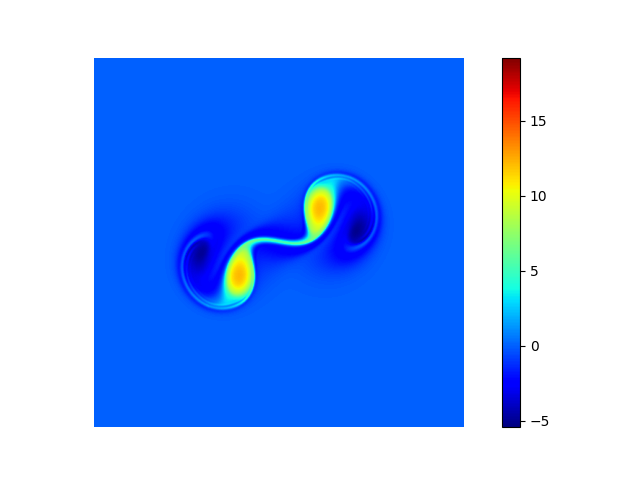}
    \negspace
    \formattedgraphics{images/validation/euler-300.png}
    \negspace
    \formattedgraphicslst{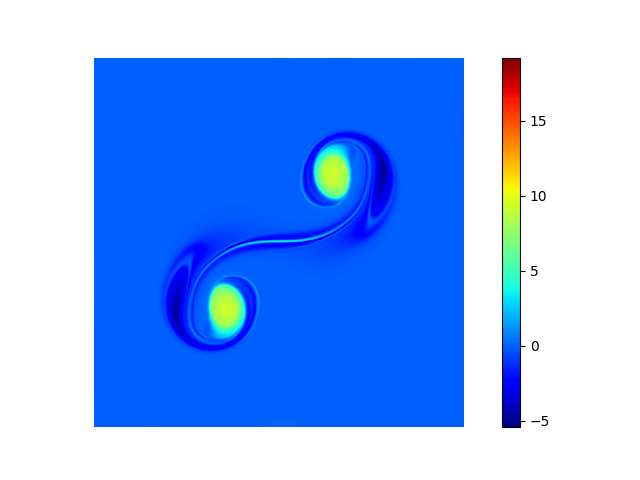}
  \end{minipage} \\[-6pt]
  \textbf{FLIP \newline $128\times 128$ with \newline $65536$ particles \newline ($1$ MB)} & 
  \begin{minipage}{0.8\textwidth}\negvspace
    \formattedgraphics{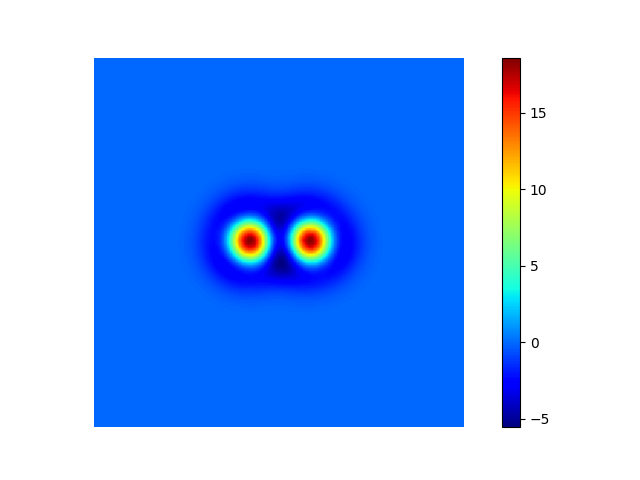}
    \negspace
    \formattedgraphics{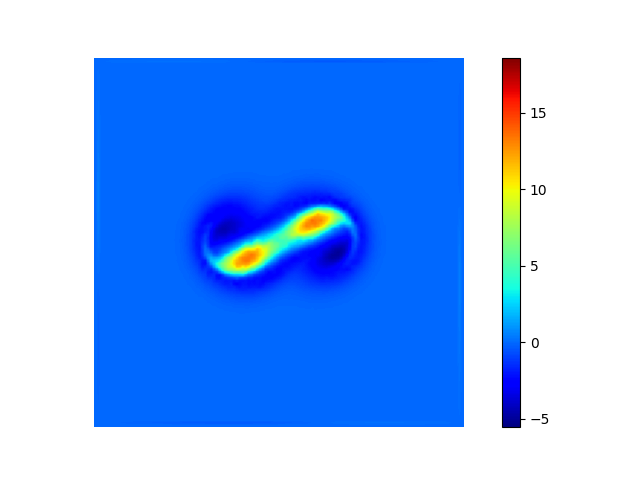}
    \negspace
    \formattedgraphics{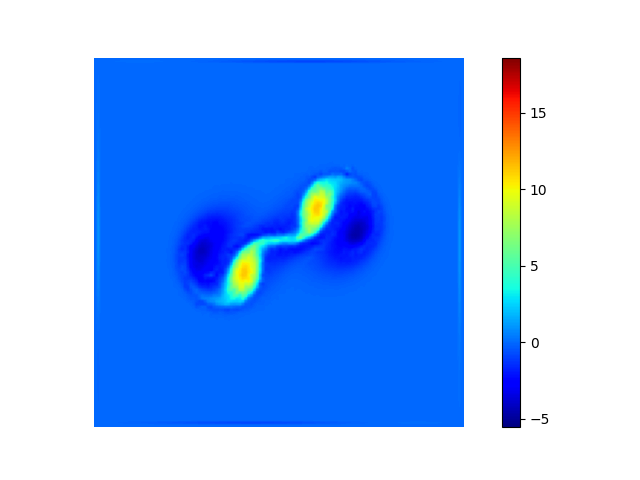}
    \negspace
    \formattedgraphics{images/validation/flip-300.png}
    \negspace
    \formattedgraphicslst{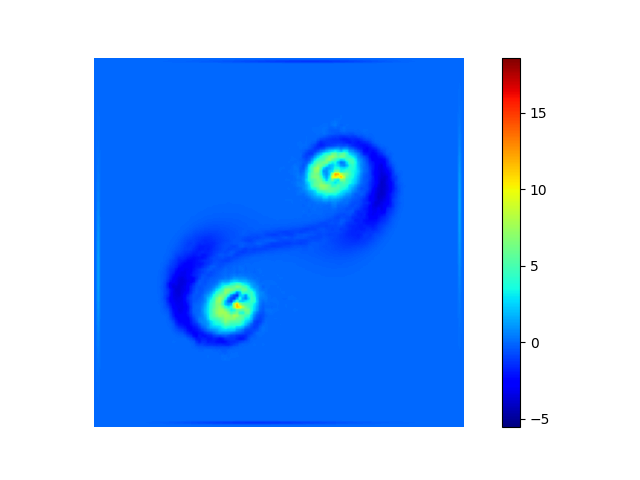}
  \end{minipage} \\[-6pt]
  \textbf{\xing{SPH \newline $288212$ particles \newline ($8.8$ MB)}} & 
  \begin{minipage}{0.8\textwidth}\negvspace
    \formattedgraphics{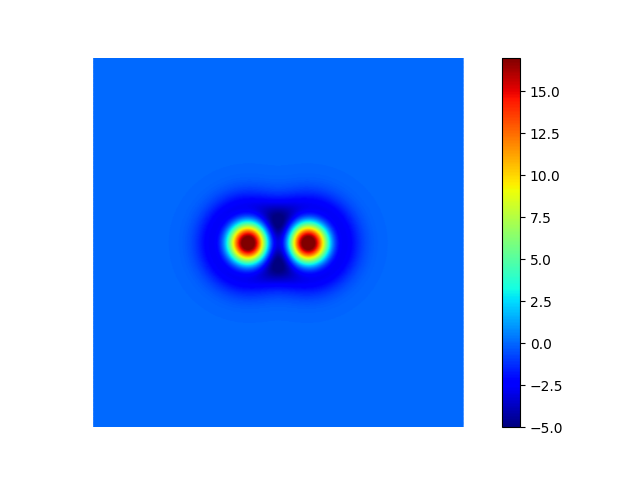}
    \negspace
    \formattedgraphics{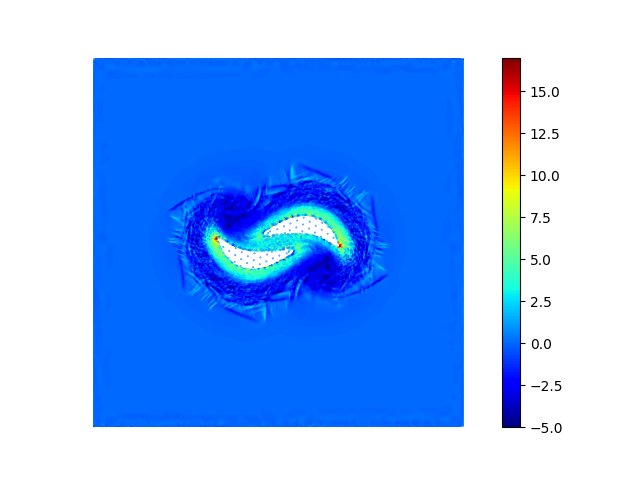}
    \negspace
    \formattedgraphics{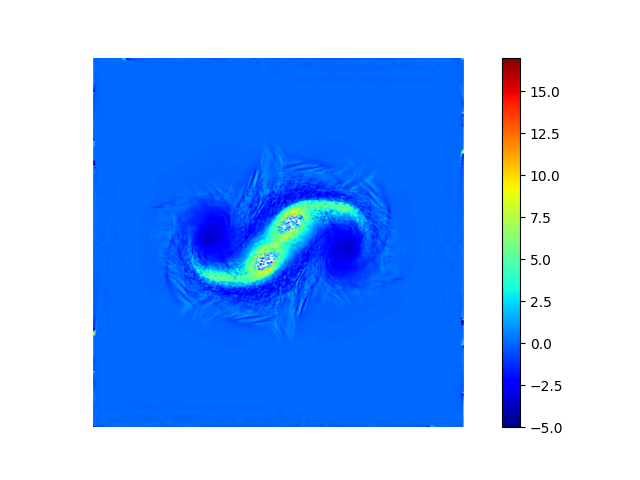}
    \negspace
    \formattedgraphics{images/validation/sph-300.png}
    \negspace
    \formattedgraphicslst{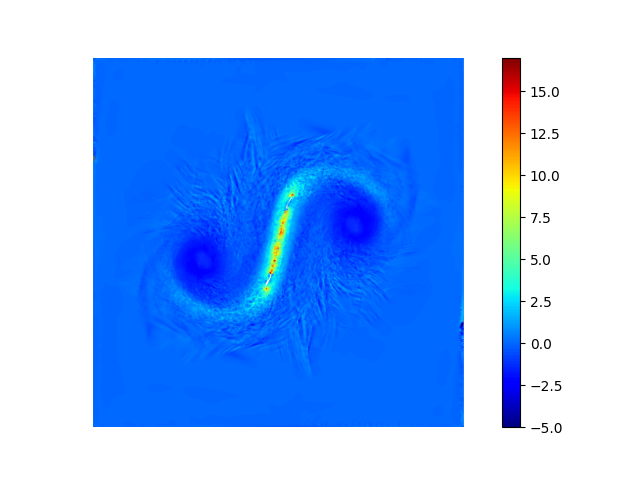}
  \end{minipage} \\[-6pt]
  \textbf{INSR \newline \cite{chen2023implicit} \newline ($32.1$ KB)} & 
  \begin{minipage}{0.8\textwidth}\negvspace
    \formattedgraphics{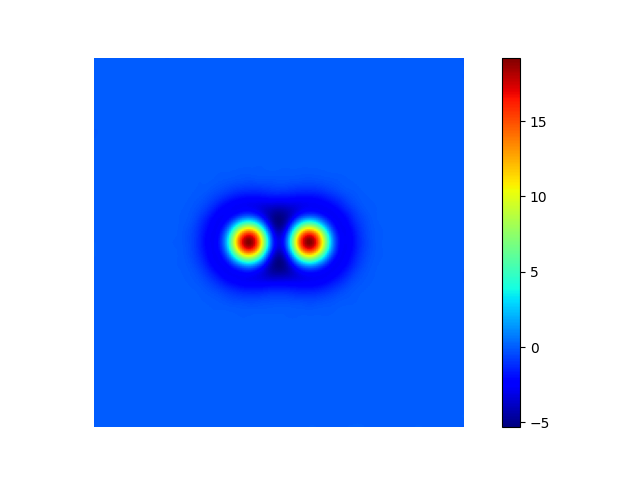}
    \negspace
    \formattedgraphics{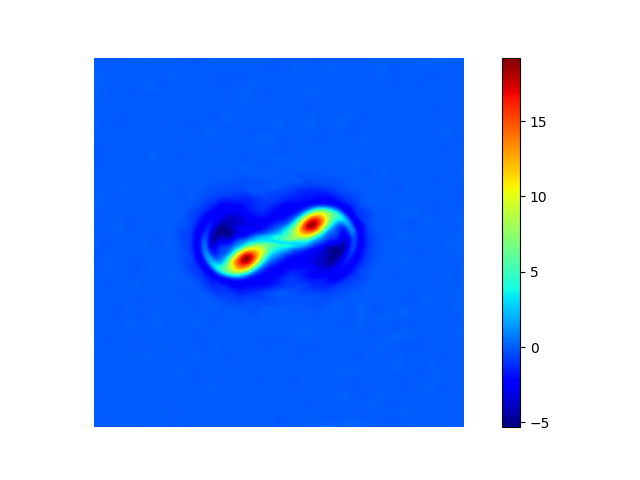}
    \negspace
    \formattedgraphics{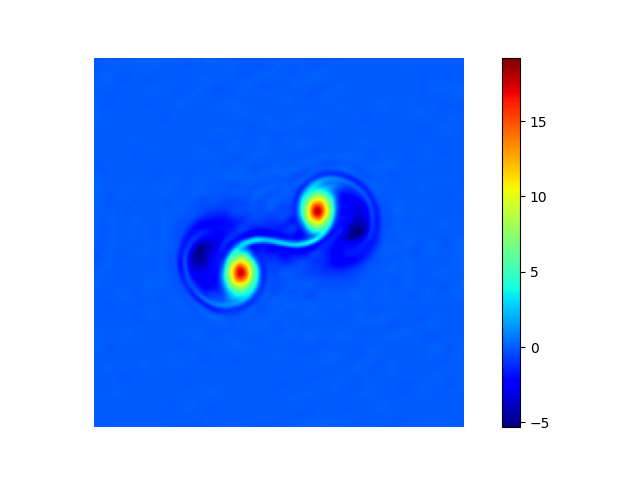}
    \negspace
    \formattedgraphics{images/validation/INSR-300.png}
    \negspace
    \formattedgraphicslst{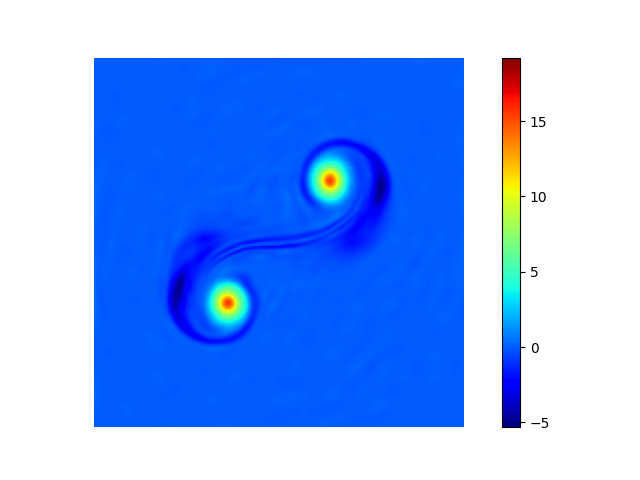}
  \end{minipage} \\[-6pt]
  \textbf{Ours \newline $5041\sim 5511$ particles \newline ($148.5$ KB)} & 
  \begin{minipage}{0.8\textwidth}\negvspace
    \formattedgraphics{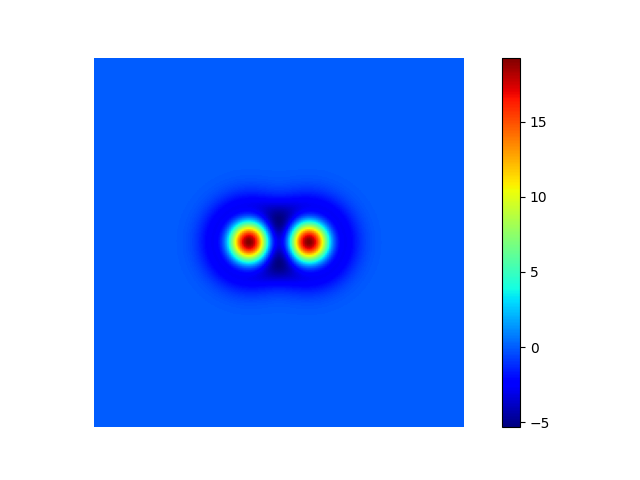}
    \negspace
    \formattedgraphics{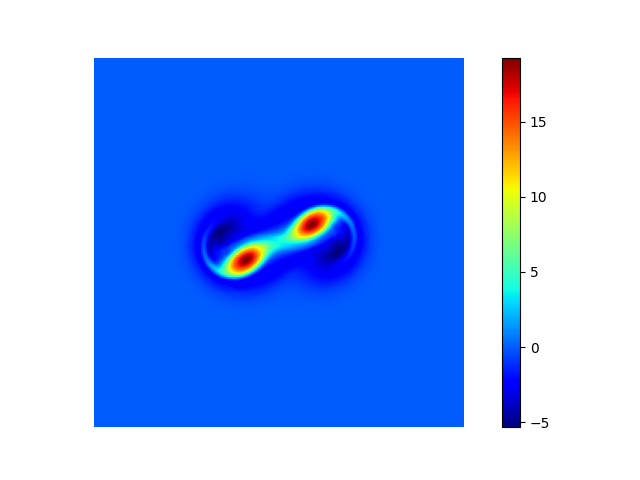}
    \negspace
    \formattedgraphics{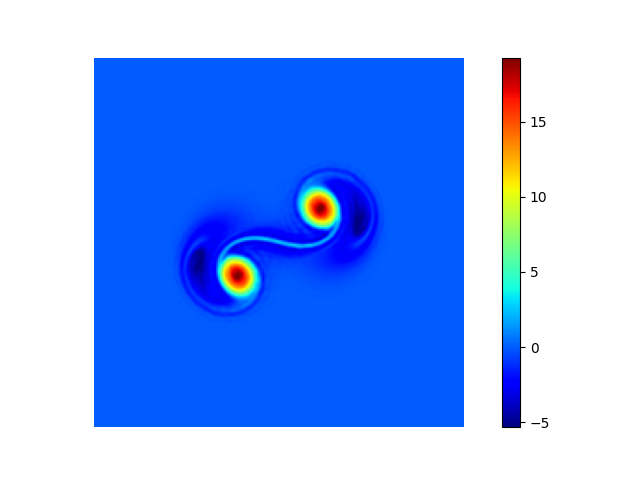}
    \negspace
    \formattedgraphics{images/validation/ours-300.png}
    \negspace
    \formattedgraphicslst{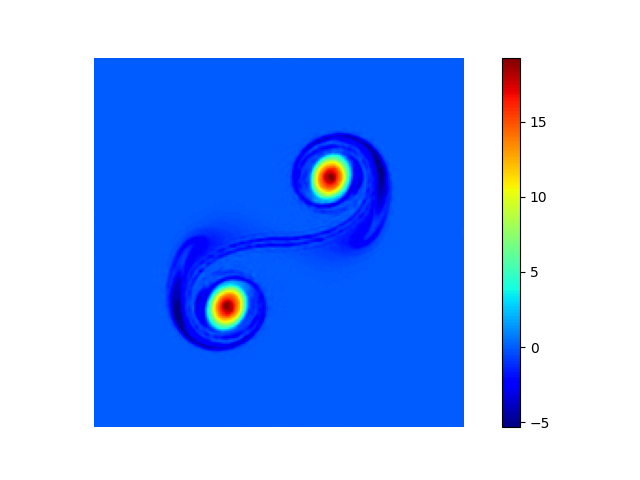}
  \end{minipage} \\[-6pt]
  \textbf{Ground truth: \newline vortex-in-cell \newline $512\times 512$ \newline ($2$ MB)} & 
  \begin{minipage}{0.8\textwidth}\negvspace
    \formattedgraphics{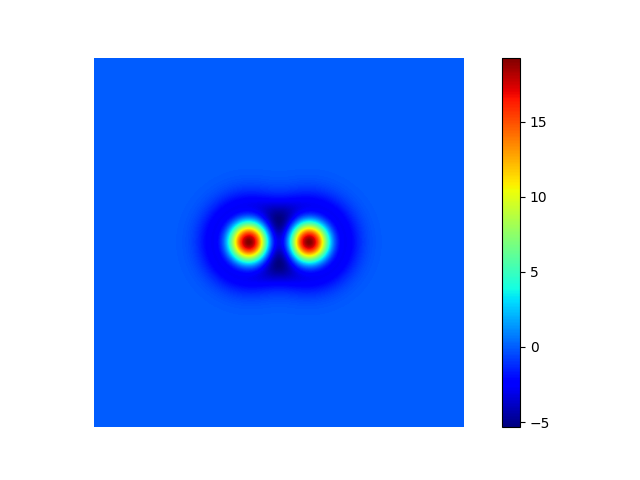}
    \negspace
    \formattedgraphics{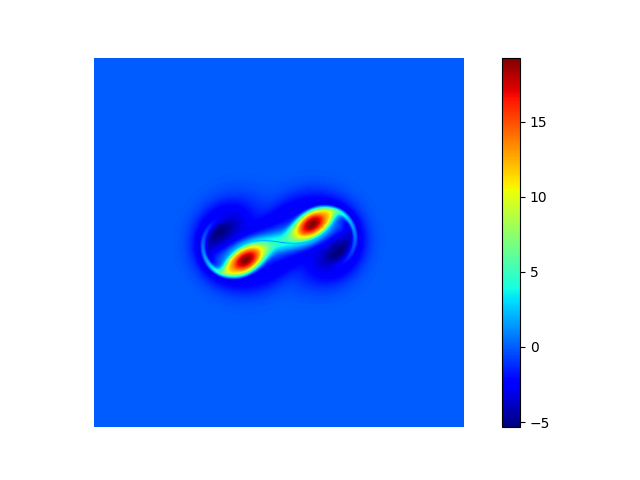}
    \negspace
    \formattedgraphics{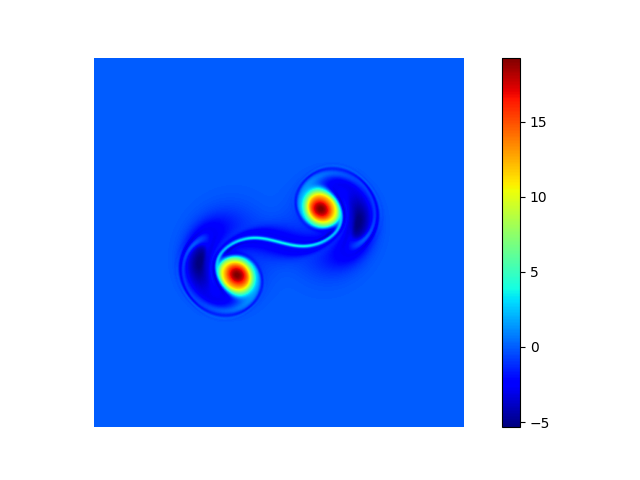}
    \negspace
    \formattedgraphics{images/validation/vortex-300.png}
    \negspace
    \formattedgraphicslst{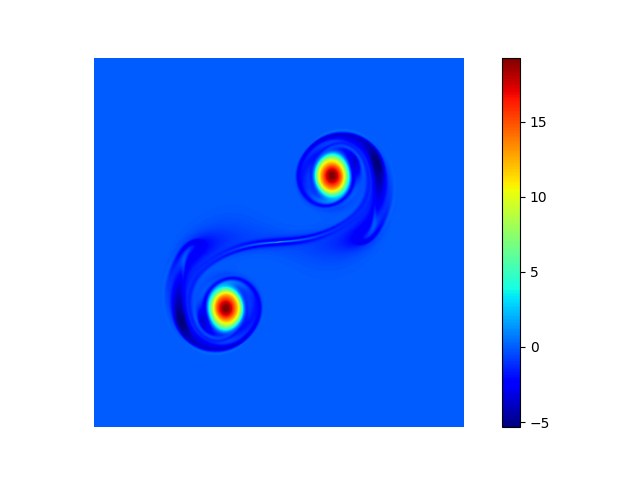}
  \end{minipage} \\[-6pt]
\end{tabular}
\caption{\footnotesize We compare our method with different traditional spatial discretizations and INSR using the Taylor vortex example. We show the average file size of the according representation of the velocity fields on the first column, enclosed by parenthesis. Images show the vorticity field at frame 0, 100, 200, 300, and 399, respectively. \xing{Note the initial frame of the SPH results is sightly different from others since the vorticity on each particle is calculated with the differential operators in the SPH convention.}}
\label{fig:comparison}
\vspace{12pt}
    \centering
    \includegraphics[width=0.84\textwidth]{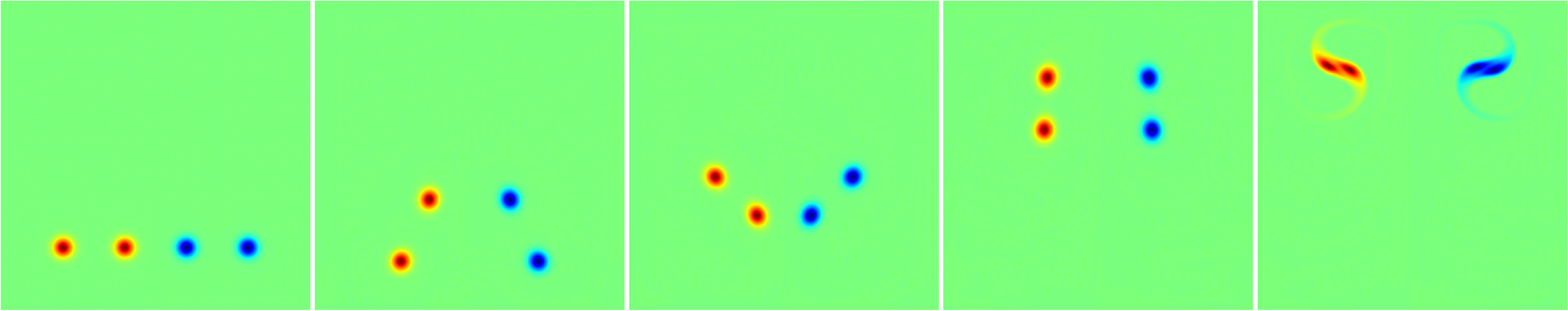}
    \vspace{-12pt}
    \captionof{figure}{\footnotesize Our simulation results on the 2D leapfrog example. The figures are showing frames 0, 165, 456, 1050 and 1500 from left to right.}
    \label{fig:leapfrog}
\end{figure*}

\paragraph{Efficacy of the GSR}
We validate the efficacy of our proposed GSR by comparing the simulation results of Taylor vortex generated by our method, traditional methods with explicit representations, and INSR as another continuous representation. As shown in Figure~\ref{fig:taylor-cmp}, the GSR preserves vorticity significantly better than all other methods, offering the best approximation to the ground truth, generated using a vortex-based 2D fluid solver. More frames are visualized in Figure~\ref{fig:comparison}.
Compared to the Eulerian method with a $512 \times 512$ MAC grid, the GSR preserves the thin structure of the Taylor vortex with significantly reduced memory. When compared to the FLIP method with a $128 \times 128$ grid and $65536$ particles, GSR demonstrates superior detail preservation and stability, using fewer than $5600$ particles, highlighting the advantages of Gaussian kernels over traditional Lagrangian particles in both accuracy and stability. \xing{Our method also show far less numerical dissipation free of artifacts caused by particle deficiency compared to the SPH method.}
While both INR and GSR represent fluid details with lightweight data structures, our method outperforms INSR by better preserving vorticity and maintaining a cleaner background due to the locality of the Gaussian kernels. Additionally, our method is more time-efficient, as shown in Table~\ref{tab:performance}, due to the high efficiency of differentiating GSR.

\paragraph{Effectiveness of Our Projection Optimization}
Our projection step demonstrates stability in both obstacle-free and boundary-driven scenarios. We evaluate the efficiency of our optimization by comparing it to other optimization-based methods, INSR and NMC, on two classic phenomena: the Taylor vortex (without obstacles) and the Karman vortex street (boundary-driven). Compared to INSR, our method introduces less numerical error, yielding more accuracy, as shown in Figure~\ref{fig:comparison}. Unlike INSR, our projection step avoids solving for the pressure field or Poisson equation, eliminating the calculation for third-order derivatives in optimization and providing a six-fold performance boost, as shown in Table~\ref{tab:performance}. Additionally, in the Karman vortex street example (Figure~\ref{fig:karman}), our method demonstrates improved stability and accuracy, effectively handling boundary conditions in a long time horizon. While NMC encounters numerical instability at the domain boundary, our method successfully generates stable vortex shedding, closely matching the reference produced by stable fluids with reflection projection.

\paragraph{Harmonic Component Validation}
\xing{Our method better handles the harmonic components without extra efforts comparing to Lagrangian vortex methods.}
\citet{Yin2023FluidCohomology} noted that the nontrivial harmonic component of the velocity field evolves over time in non-simply connected fluid domains. Figure~\ref{fig:vortices_pass} illustrates a pair of vortices passing through a gap between two spherical obstacles with a no-slip boundary, where correctly modeling the harmonic component is key for successful traversal.
\xing{We apply the Kirchoff point vortex dynamics and Kelvin's method of reflection for handling circular obstacles in the simulation by vortex particles, while maintaining conserved harmonic components using the same way as \cite{Yin2023FluidCohomology}. Unlike the vortex particles, our method successfully models the dynamics of the pair of vortices passing through the gap, indicating proper treatment of the harmonic components.}
However, our solution cannot strictly guarantee the satisfaction of the boundary condition or the divergence-free constraint, which we will discuss further in the supplementary material.

\subsection{More Examples}
\paragraph{Leapfrog 2D}
Four vortices are placed at the bottom of the domain, with the left two vortices spin counterclockwise (i.e. with positive vorticity) and the right two spin clockwise (i.e. with negative vorticity). As shown in Figure~\ref{fig:leapfrog}, our solver is able to accurately simulate the alternating forward dynamics of two pairs of vortices and their eventual merging into a single vortex.

\paragraph{Leapfrog 3D}
Figure~\ref{fig:leapfrog_3d} shows an example initialized with two parallel vortex rings facing the same direction. As they move forward, the vortex rings pass through each other, marking one leap. Our method successfully maintains the ring shapes after 3 leaps.
\xing{However, numerical error is accumulated throughout the simulation, which is then converted by the physics-based optimization into low-divergence velocity components, resulting in small vortex-ring artifacts near the end of our 3D simulations. Reducing the accumulated error is worth future exploration.}

\paragraph{Ring Collide}
In this example, two parallel vortex rings face each other at the initial frame. As shown in Figure~\ref{fig:ring_collide}, the rings expand as they approach each other, followed by shredding into many small vortex rings upon collision.
Figure~\ref{fig:ring_collide-density} shows the rendering of passive smoke rings advected by the velocity field. We also compare it with a real-world recording in our video.

\paragraph{Smoking Bunny}
To demonstrate that our method is capable of handling boundary with a complex geometry, we release two vortex rings towards the face of the Stanford Bunny. As shown in Figure~\ref{fig:smoking_bunny}, the vortex rings deforms and breaks down as they hit the uneven surface, immersing the bunny in a foggy environment.

\begin{figure}[htbp]\small
\centering
\begin{subfigure}{\linewidth}
        \centering
        \includegraphics[trim=300 300 500 340,clip,width=\linewidth]{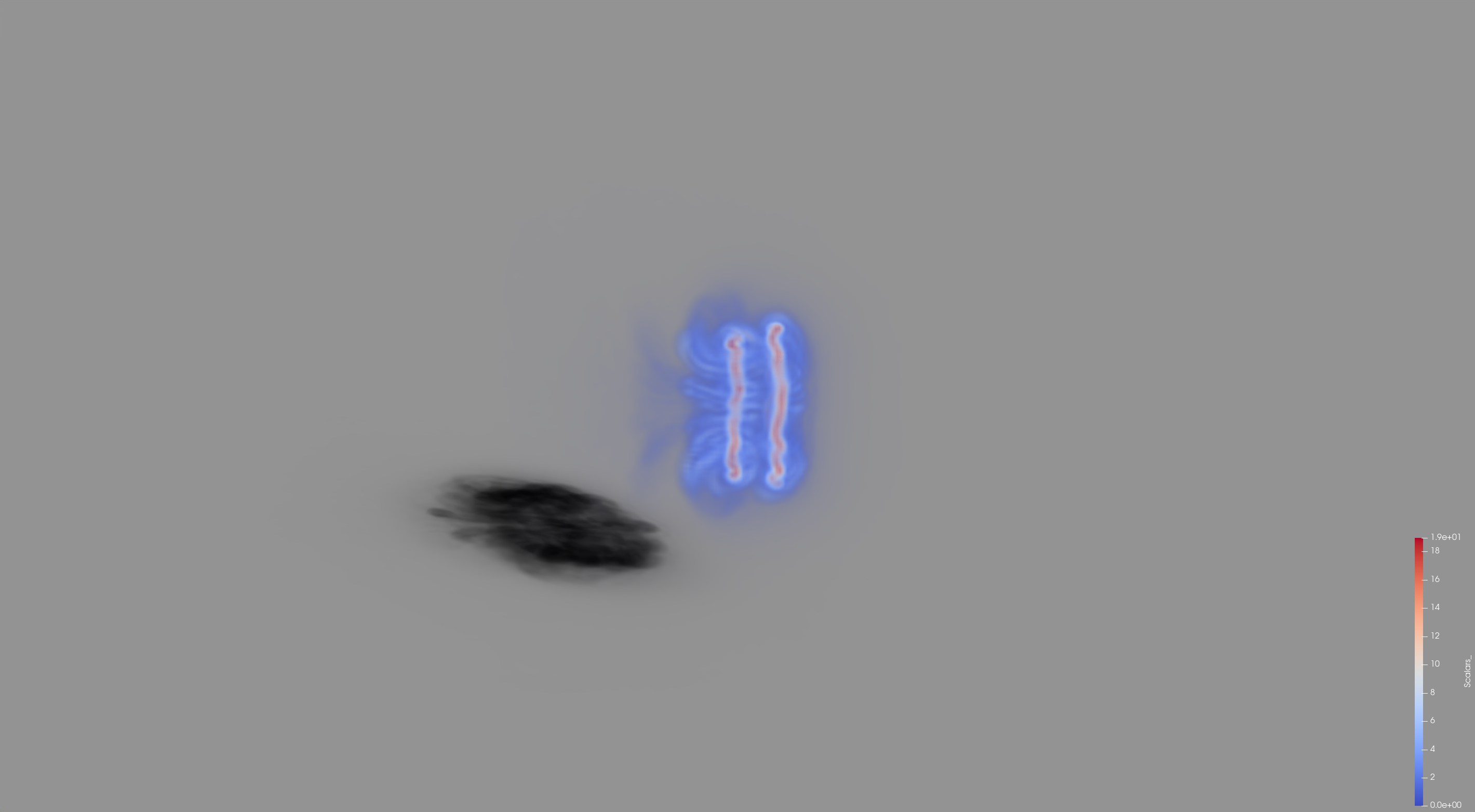}\vspace{-6pt}
        \caption{3D Leapfrog}
\end{subfigure}\\
 \begin{subfigure}{0.485\linewidth}
        \centering
        \includegraphics[trim=0 0 0 0,clip,width=\linewidth]{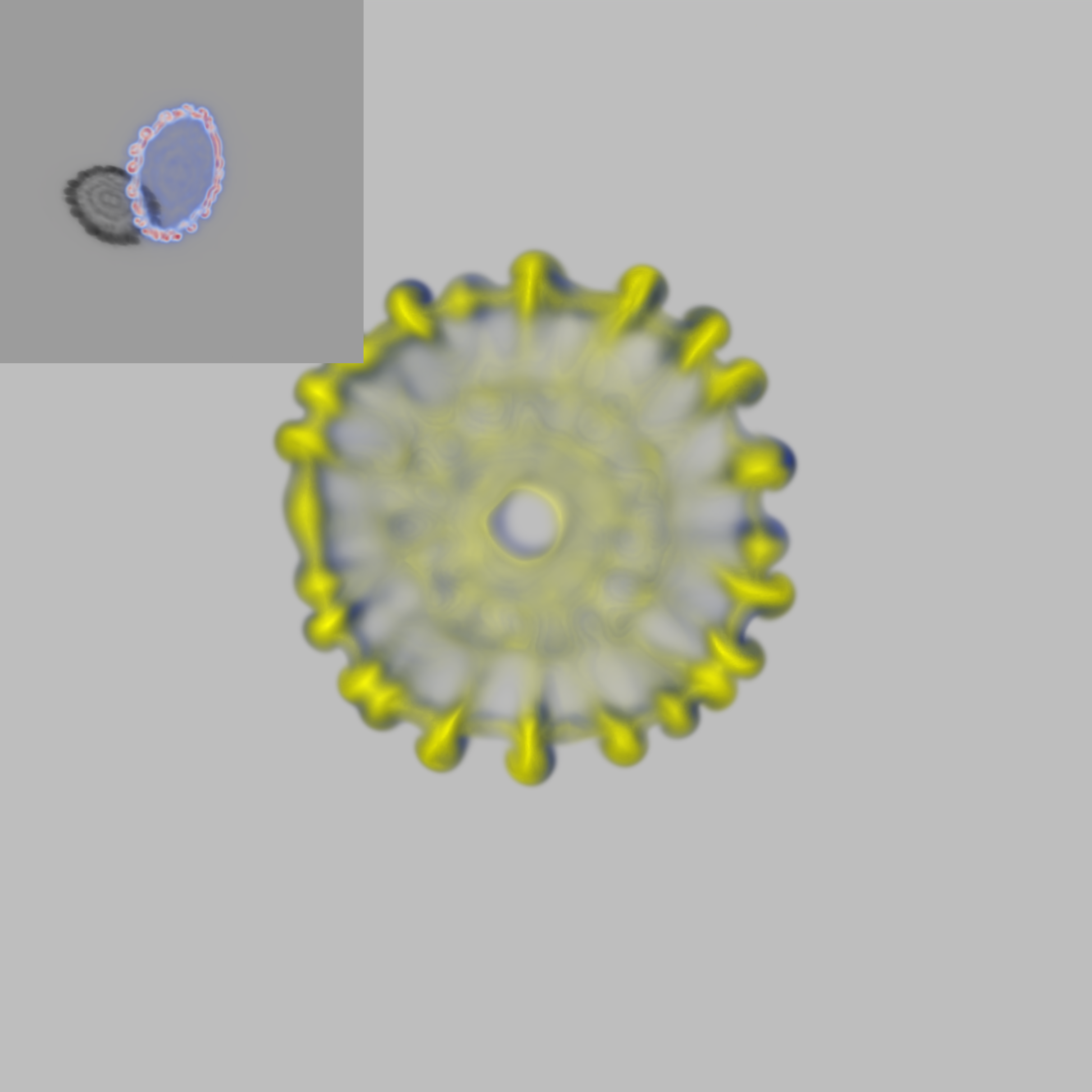}\vspace{-6pt}
        \caption{Ring Collide}
\end{subfigure}
\hfill
 \begin{subfigure}{0.504\linewidth}
        \centering
        \includegraphics[trim=8 2 10 0,clip,width=\linewidth]{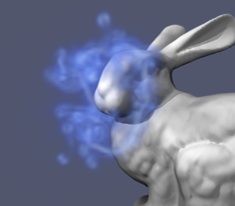}\vspace{-6pt}
        \caption{Bunny}
\end{subfigure}\vspace{-12pt}
\caption{\small We test our method in complex scenes, demonstrating the advantages of stability and spatial details.}
\end{figure}

\subsection{Ablation Tests}
\label{sec:Ablation}

\begin{figure*}[htbp]
    \centering
    \newcommand{\formattedgraphicsA}[1]{\includegraphics[trim=700 300 900 250,clip,width=0.2\textwidth]{#1}}
    \newcommand{\negspaceA}[0]{\hspace{-3pt}}
    \formattedgraphicsA{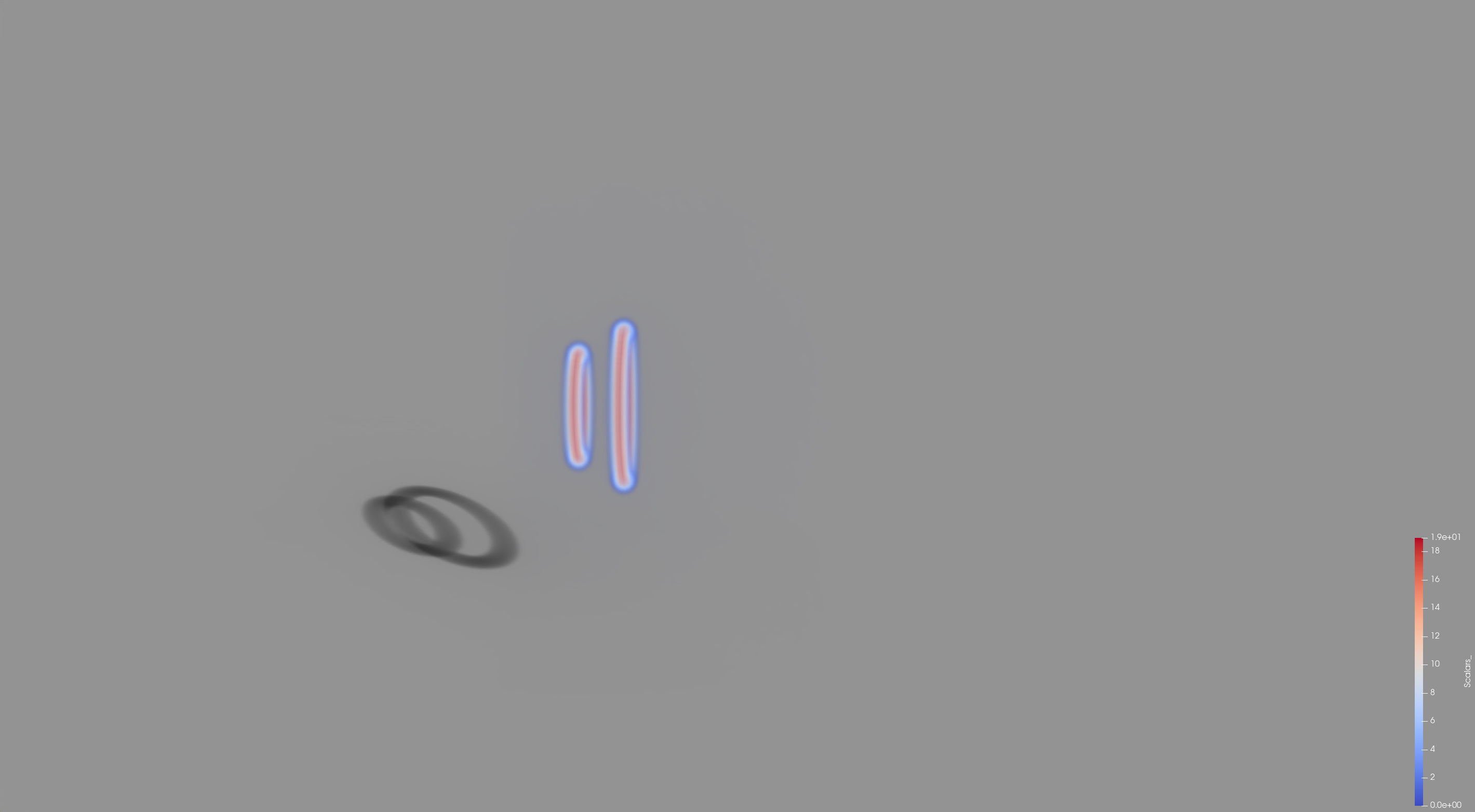}
    \negspaceA
    \formattedgraphicsA{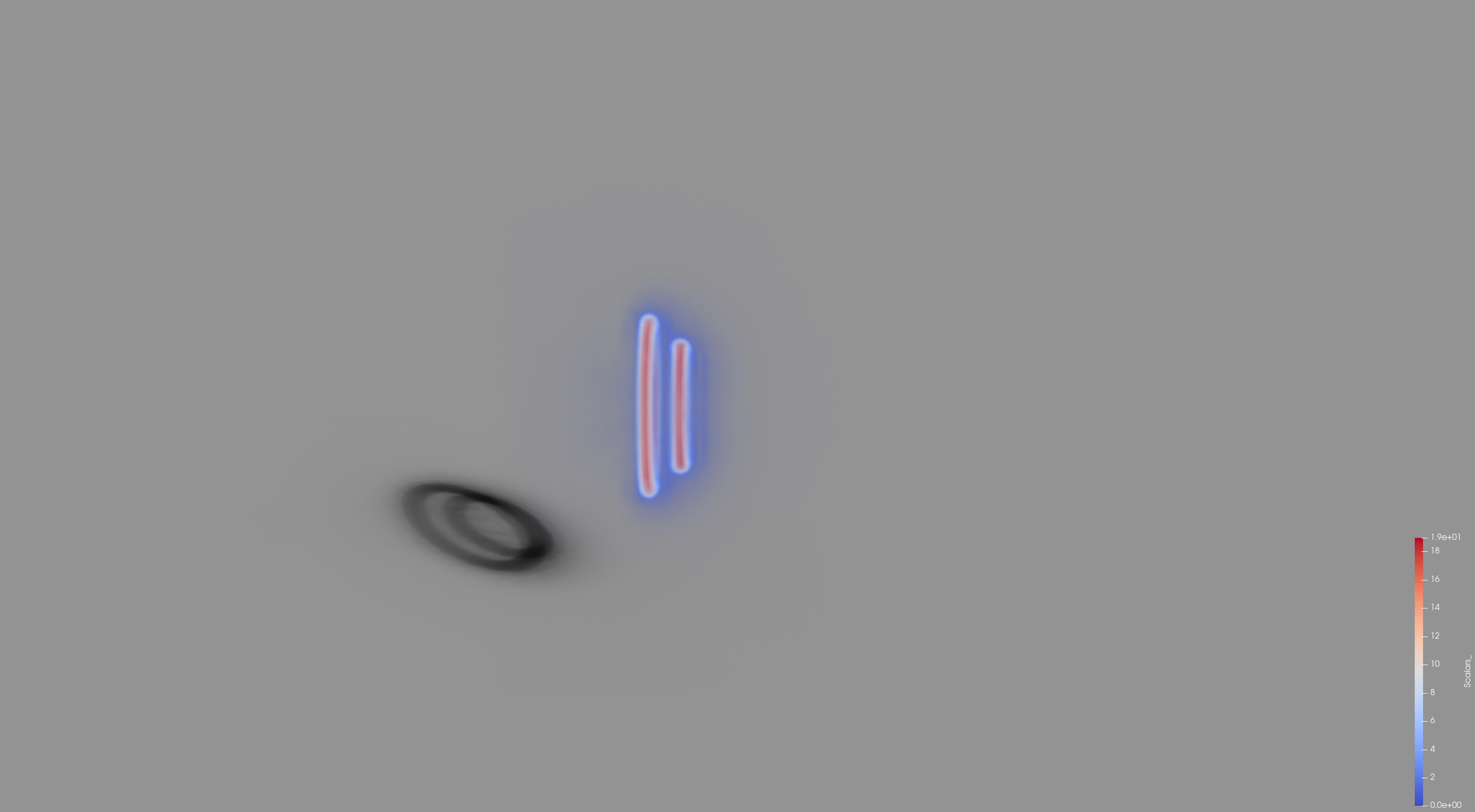}
    \negspaceA
    \formattedgraphicsA{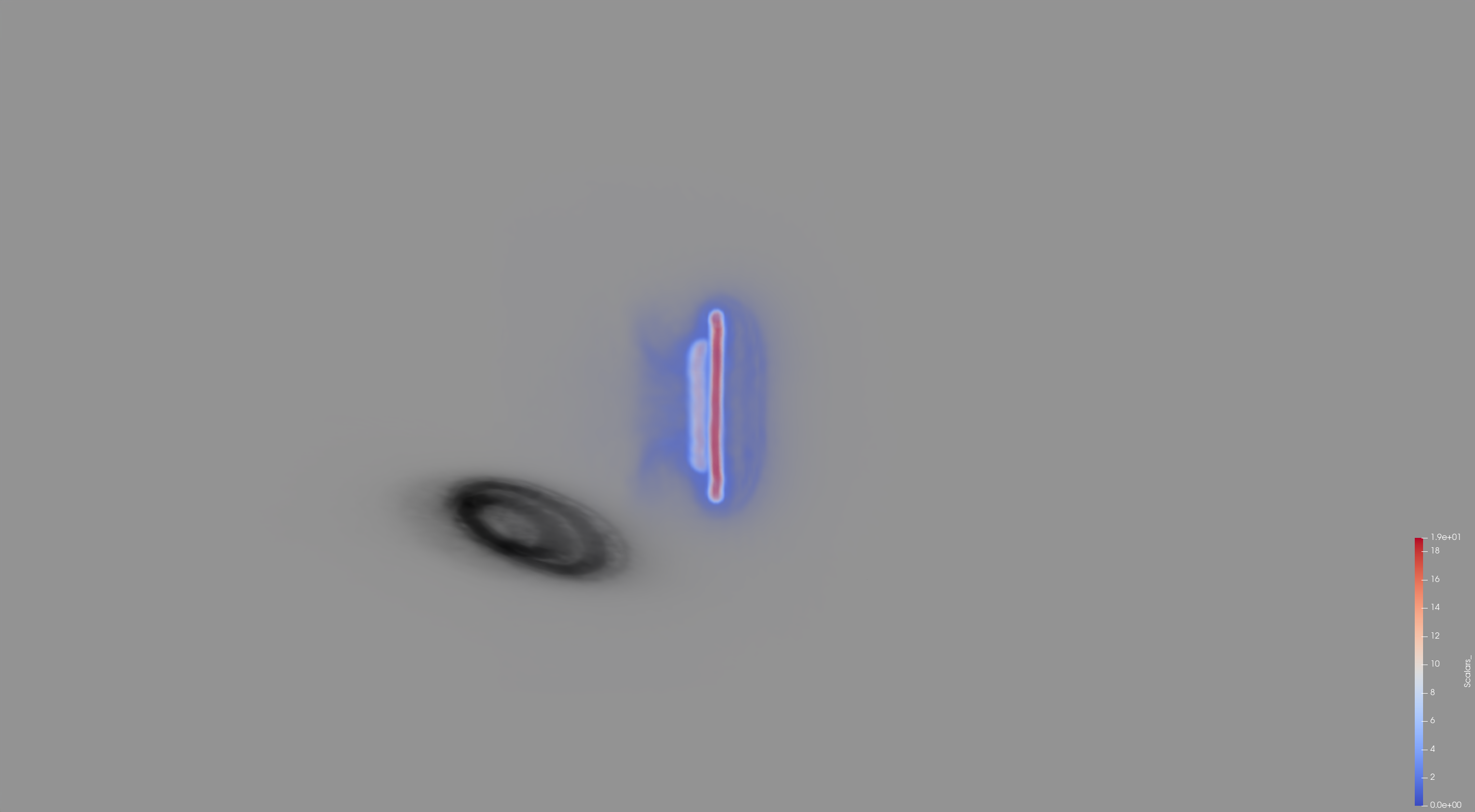}
    \negspaceA
    \formattedgraphicsA{images/leapfrog_3d/leapfrog.0548.png}
    \negspaceA
    \formattedgraphicsA{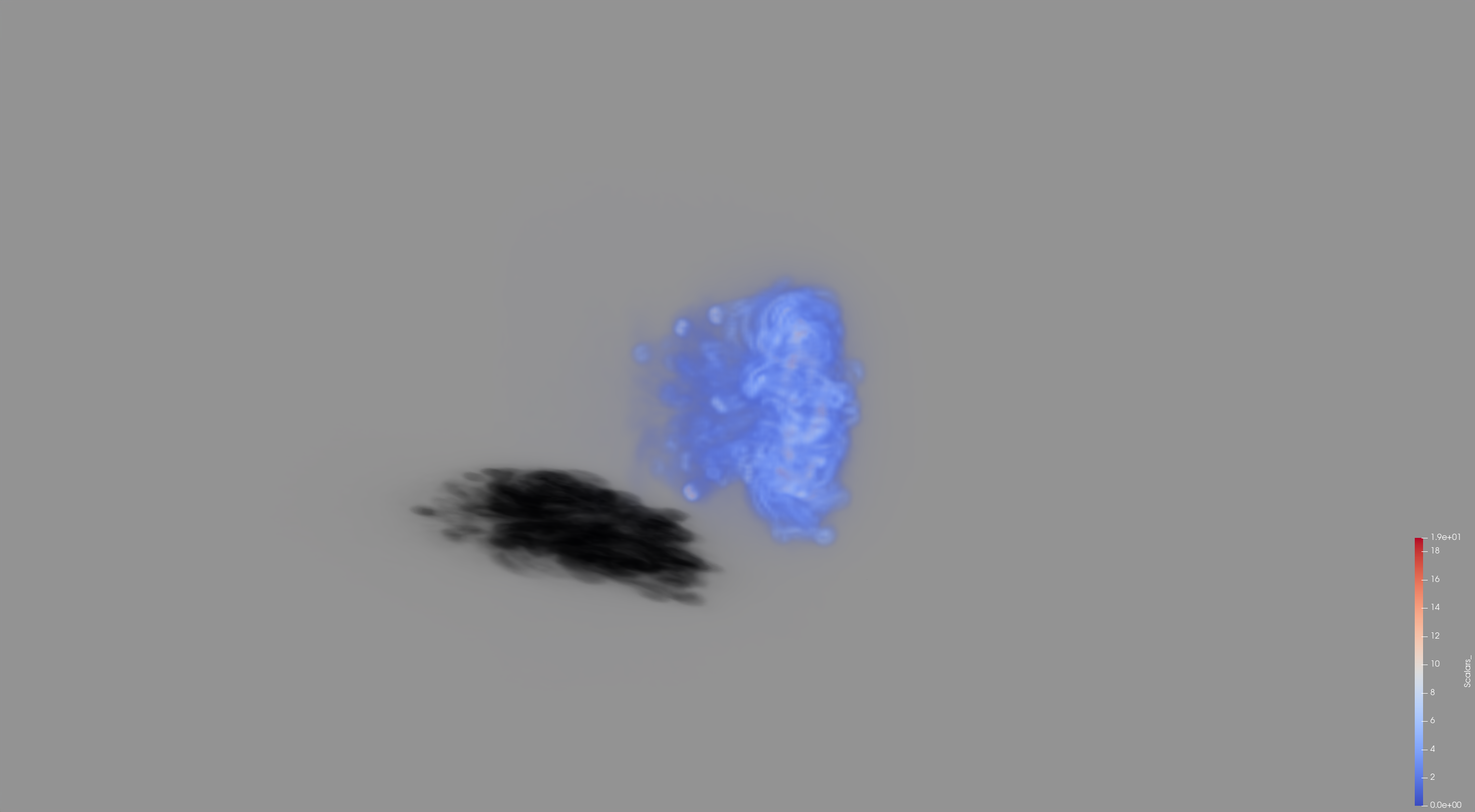}
    \vspace{-12pt}
    \caption{Our simulation results on the 3D leapfrog example. The figures are frames 0, 148, 332, 548, and 860 from left to right.}
    \label{fig:leapfrog_3d}
    \vspace{12pt}
    
    \newcommand{\formattedgraphicsB}[1]{\includegraphics[trim=0 0 0 0,clip,width=0.2\textwidth]{#1}}
    \newcommand{\negspaceB}[0]{\hspace{-2pt}}
    \formattedgraphicsB{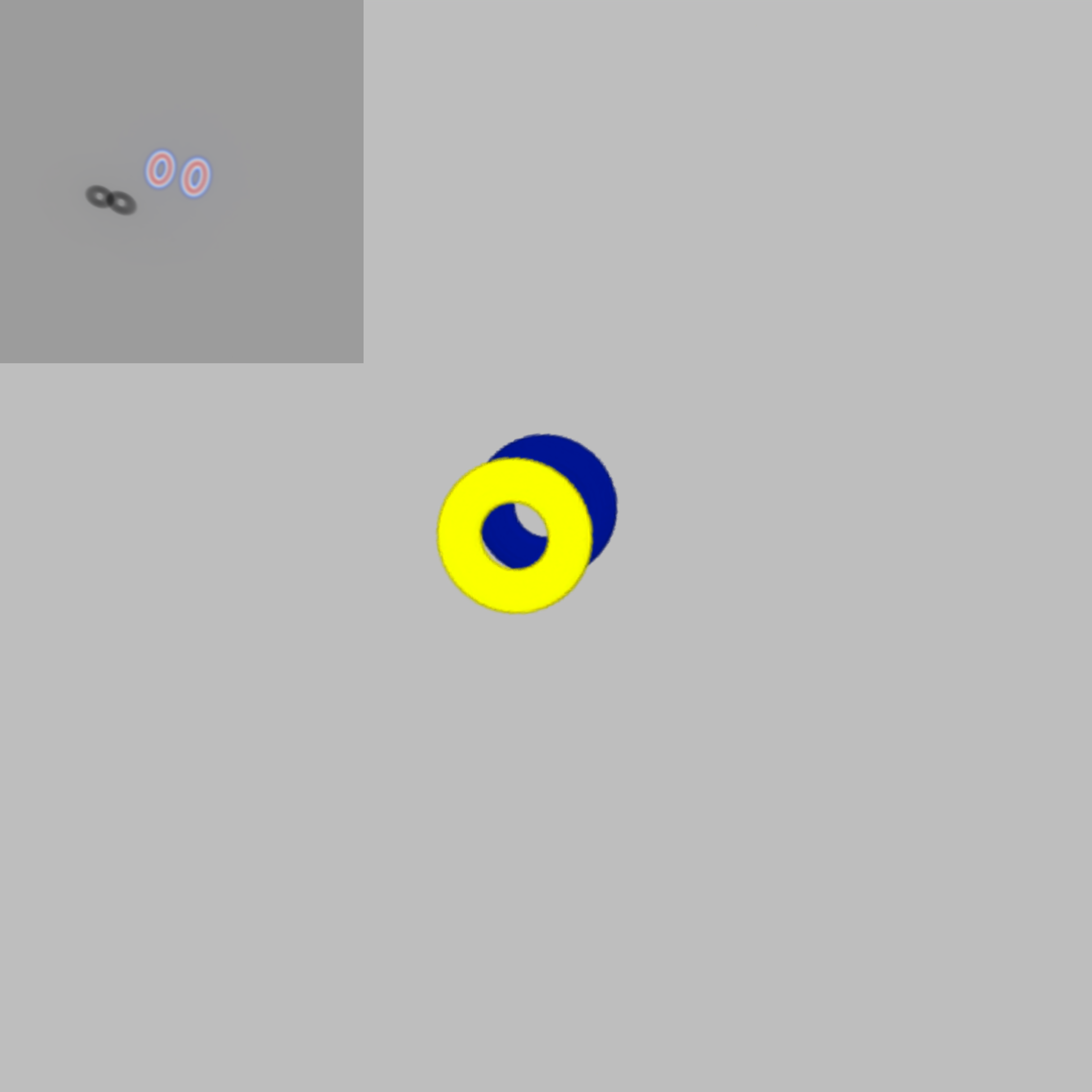}
    \negspaceB
    \formattedgraphicsB{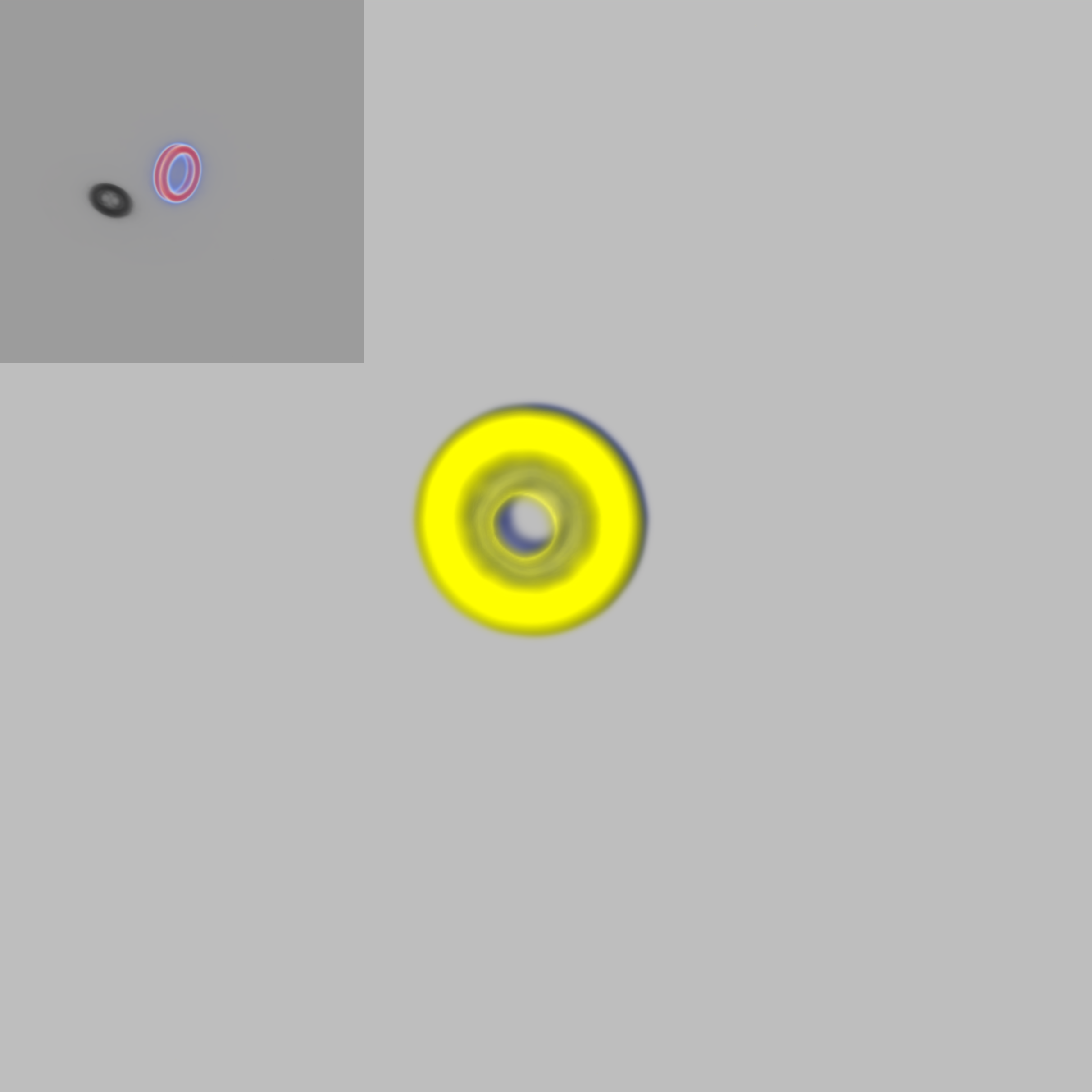}
    \negspaceB
    \formattedgraphicsB{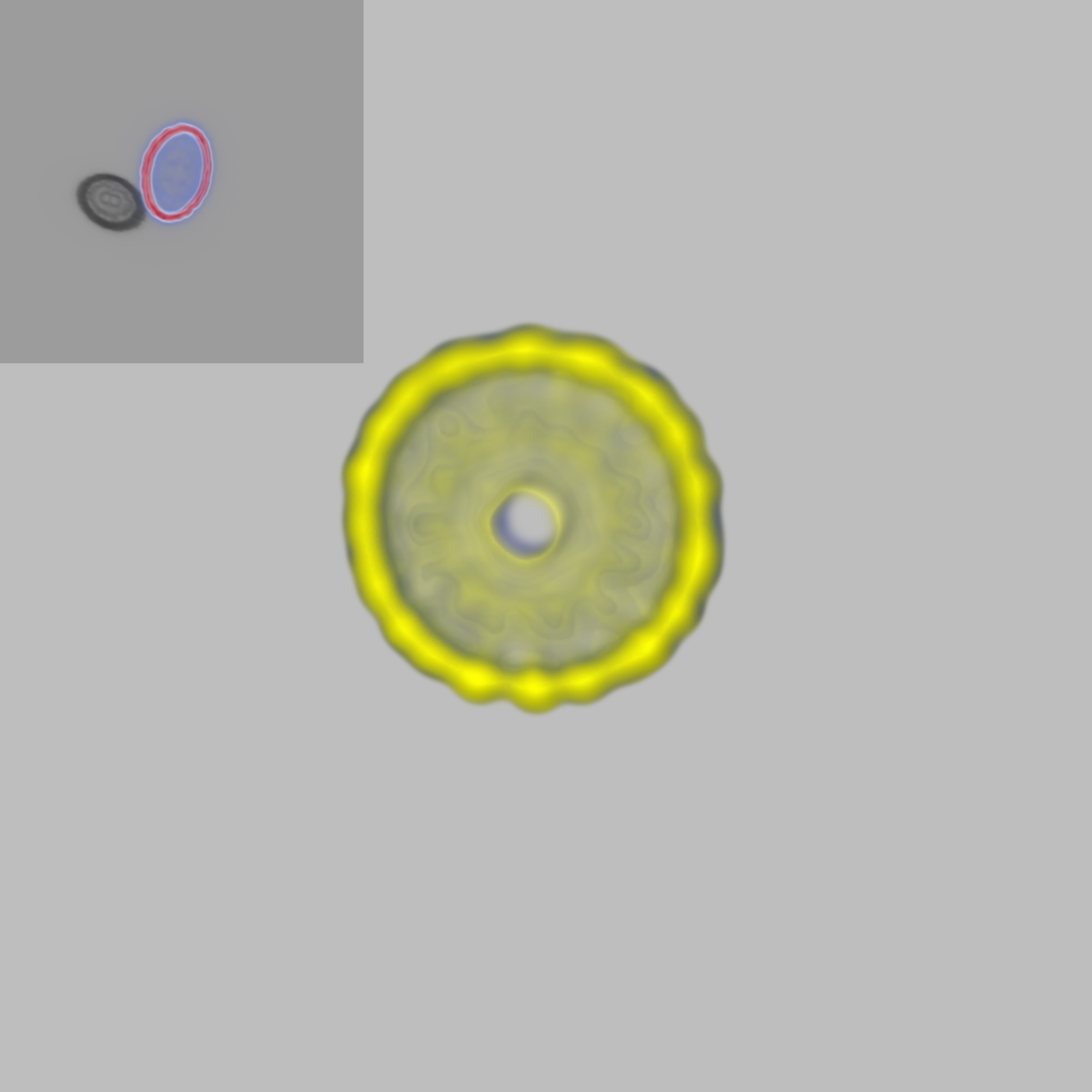}
    \negspaceB
    \formattedgraphicsB{images/ring_collide/ring_collide-density-178.png}
    \negspaceB
    \formattedgraphicsB{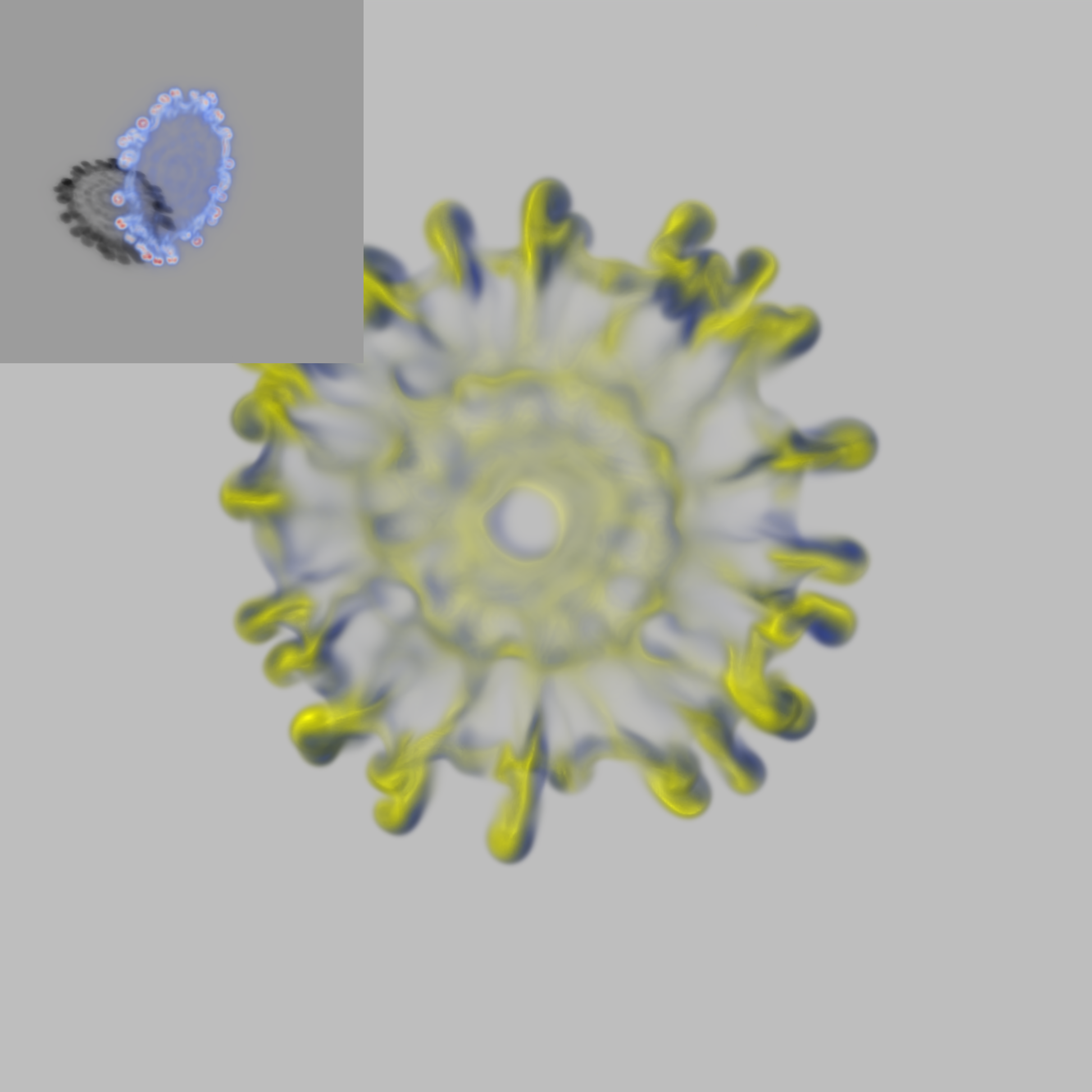}\negspaceB
    \vspace{-12pt}
    \captionof{figure}{Passive field advected by the ring collide example at frame 0, 74, 123, 178, and 242. Thumbnails of the vorticity view are placed on the top-left.}
    \label{fig:ring_collide-density}
    \vspace{12pt}
    
    \newcommand{\formattedgraphics}[1]{\includegraphics[trim=0 0 0 0,clip,width=0.2\textwidth]{#1}}
    \newcommand{\negspace}[0]{\hspace{-2pt}}
    \formattedgraphics{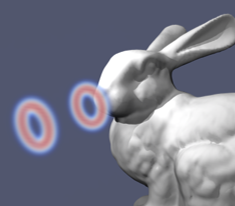}
    \negspace
    \formattedgraphics{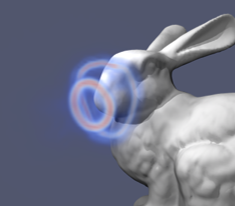}
    \negspace
    \formattedgraphics{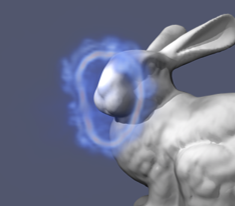}
    \negspace
    \formattedgraphics{images/smoking_bunny/bunny-300.png}
    \negspace
    \formattedgraphics{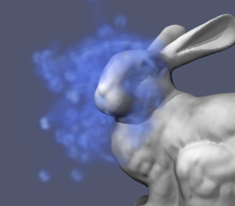}\negspace
    \vspace{-12pt}
    \captionof{figure}{Vorticity magnitude of the smoking bunny example at frame 0, 95, 188, 300, and 399 from left to right.}
    \label{fig:smoking_bunny}
    \vspace{12pt}
    
    \begin{minipage}{0.45\textwidth}
    \centering
    \begin{subfigure}{0.48\columnwidth}
        \centering
        \includegraphics[width=\textwidth]{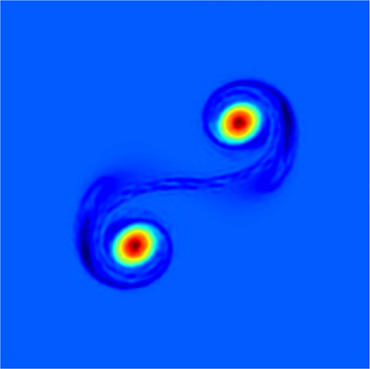}   \vspace{-18pt}
        \caption{\footnotesize Without particle splitting}
        \label{fig:wo-split}
    \end{subfigure}
    \hfill
    \begin{subfigure}{0.48\columnwidth}
        \centering
        \includegraphics[width=\textwidth]{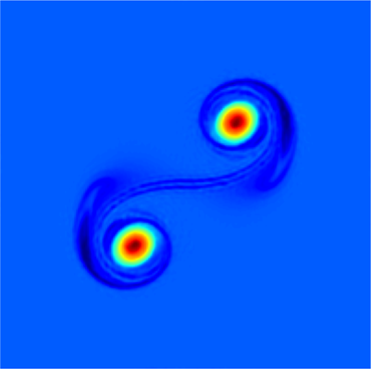}   \vspace{-18pt}
        \caption{\footnotesize Full method}
        \label{fig:with-split}
    \end{subfigure}
    \vspace{-12pt}
    \captionof{figure}{\footnotesize Ablation test on particle splitting. We show vorticity fields at frame 380.}
    \label{fig:ablation-split}
    \vspace{6pt}
    \centering
    \begin{subfigure}{0.48\columnwidth}
        \centering
        \includegraphics[trim=60 40 140 60,clip,width=\textwidth]{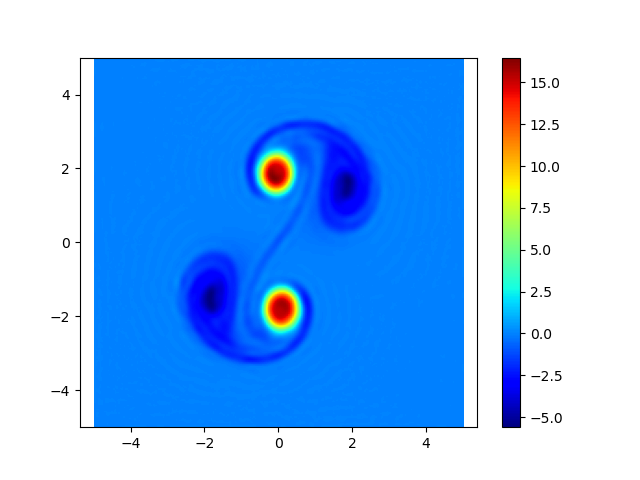}  \vspace{-18pt}  
        \caption{\footnotesize Without gradient projection}
        \label{fig:wo-grad-proj}
    \end{subfigure}
    \hfill
    \begin{subfigure}{0.48\columnwidth}
        \centering
        \includegraphics[trim=60 40 140 60,clip,width=\textwidth]{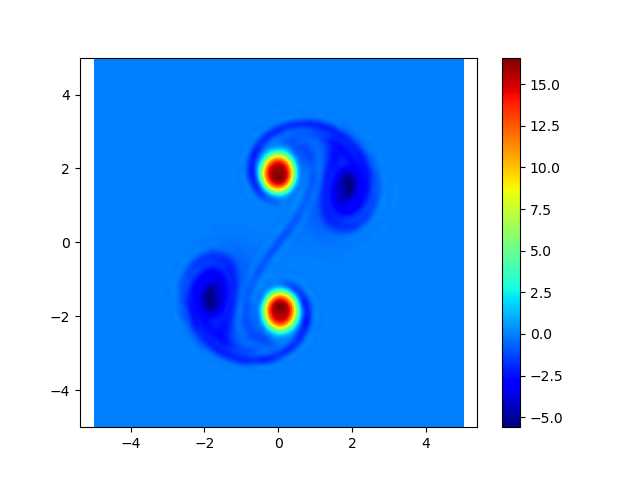} \vspace{-18pt}
        \caption{\footnotesize Full method}
        \label{fig:with-grad-proj}
    \end{subfigure}
    \vspace{-12pt}
    \captionof{figure}{\footnotesize Ablation test on the gradient projection technique. We show vorticity fields at frame 36 with a time step of 0.1 seconds.}
    \label{fig:ablation-grad-proj}
  \end{minipage}
  \hfill
  \begin{minipage}{0.53\textwidth}
    \centering
    \begin{subfigure}{.49\columnwidth}
        \centering
        \includegraphics[width=\textwidth]{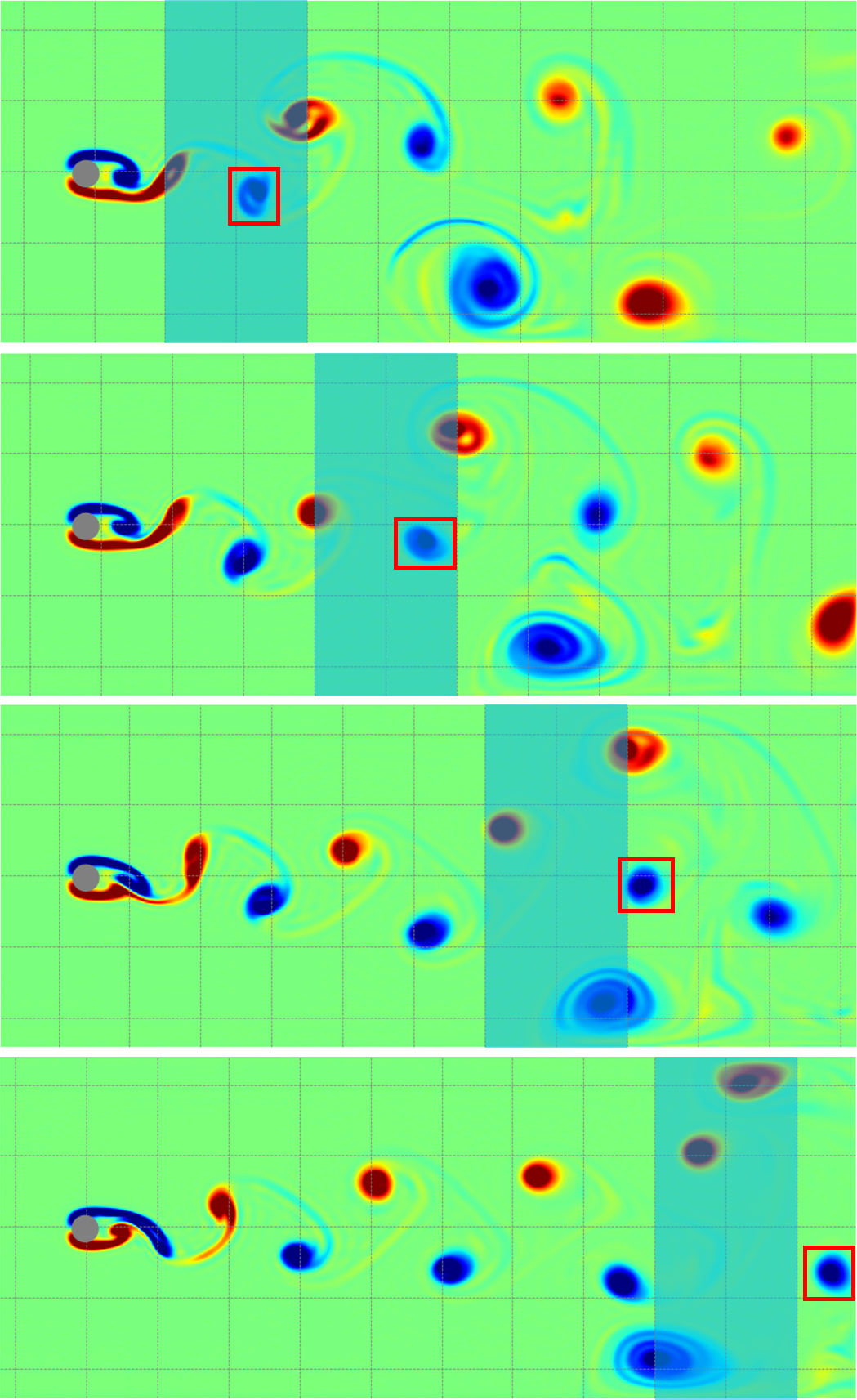}    \vspace{-12pt}
        \caption{Without \xing{initial guess}}
        \label{fig:karman-noadvect}
    \end{subfigure}
    \hfill
    \begin{subfigure}{.49\columnwidth}
        \centering
        \includegraphics[width=\textwidth]{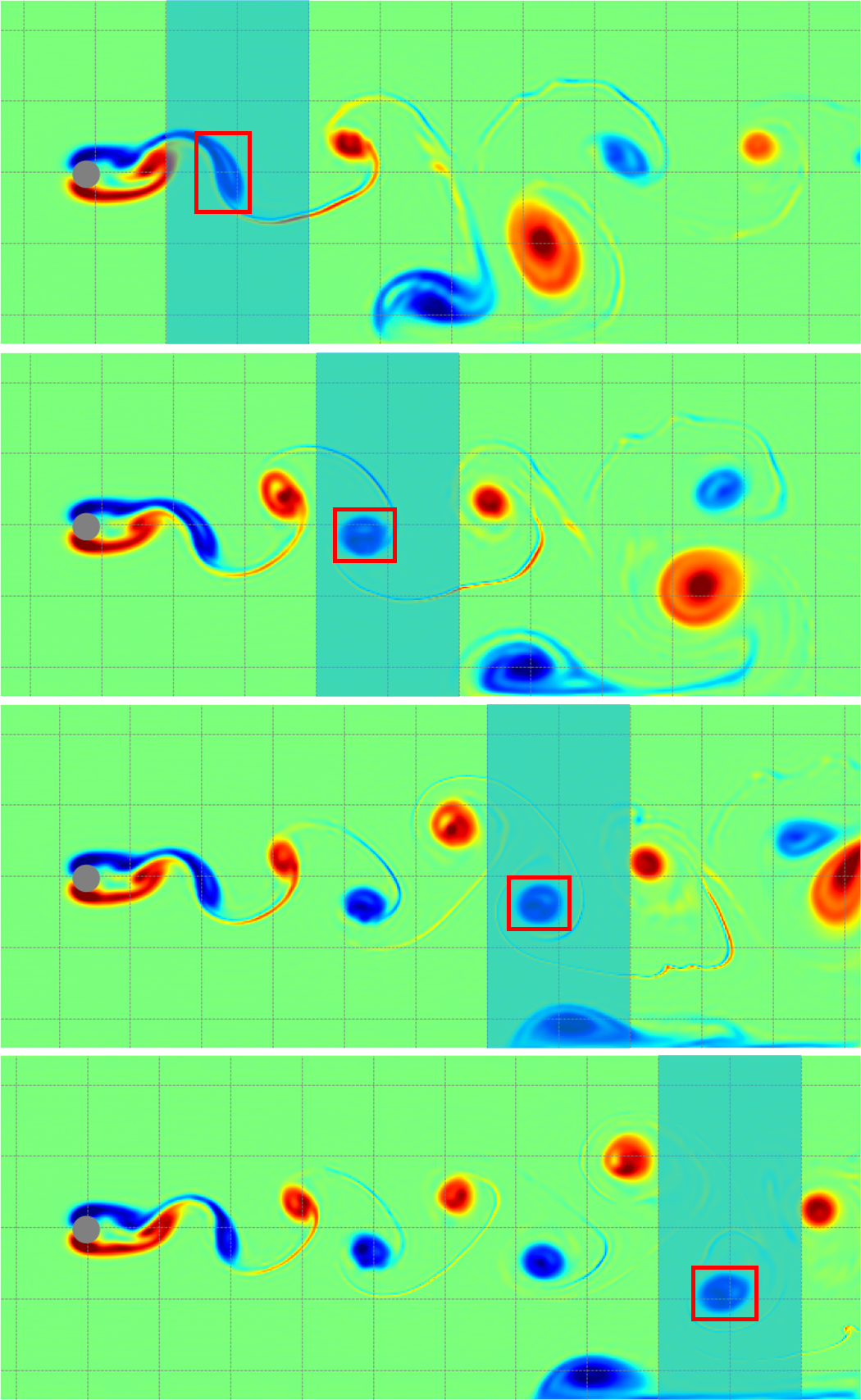}  \vspace{-12pt}
        \caption{Full method}
        \label{fig:karman-full}
    \end{subfigure}
    \vspace{-12pt}
    \captionof{figure}{\footnotesize Ablation test on the advection\xing{-based initial guess}. The images from top to bottom are simulation results at frames 127, 150, 174 and 198, respectively. The background grid in gray dashed line is moving forward at the inflow speed. We track one vortex (marked by the red boxes) to illustrate the flow rate of the simulation w.r.t. the speed of the grid.}
    \label{fig:karman-ablation}
  \end{minipage}
\end{figure*}

\paragraph{Particle Splitting}
Figure~\ref{fig:ablation-split} compares the method without particle splitting with the full method. The former fails to preserve the thin filament of the Taylor vortex due to limited expressiveness. Our reseeding strategy effectively addresses this without introducing significant overhead, as shown by the particle count in Table~\ref{tab:performance}.
\xing{Meanwhile, we do not observe any change in convergence rate of the following optimization caused by the reseeding process.}

\paragraph{Gradient Projection}
Without the gradient projection technique outlined in Section~\ref{sec:grad-proj}, directly combining the losses can be difficult to optimize. The gradient of the divergence loss may increase the vorticity loss, leading to ripple artifacts in the vorticity field, as shown in Figure~\ref{fig:ablation-grad-proj}a. However, by applying the gradient projection, we achieve a clean result, as seen in Figure~\ref{fig:ablation-grad-proj}b.


\paragraph{Advection-Based Initial Guess}
\xing{The initial guess improves both solver accuracy and convergence speed of our method.}
Figure~\ref{fig:karman-ablation} compares our method with and without \xing{the initial guess, in which case we directly optimize the GSR from its configuration at the end of the last time step}. For an intuitive comparison, we set up a background grid moving at the inflow speed with two columns marked in a darker color. 
In Figure~\ref{fig:karman-ablation}a, the flow rate is overly rapid, as evidenced by the vortex originally inside the darker zone moving out of it. In contrast, the flow rate in Figure~\ref{fig:karman-ablation}b closely matches the inflow. Additionally, without advection, the projection step requires an average of $4661.3$ iterations, compared to $3902.5$ iterations with advection.


\subsection{Performance}

We run all experiments on a NVIDIA GeForce RTX 4090 GPU. Table~\ref{tab:performance} shows the time cost and memory usage for all examples, along with a comparison to INR-based methods. Our method exhibits comparable time and memory usage, while delivering superior quality and stability.
\xing{Note the running times shown in Table~\ref{tab:performance} are all much longer than those used by traditional Lagrangian and Eulerian methods, since both GSR and INR-based methods require costly first-order optimizations.}

\section{Conclusion}
We have presented a novel grid-free fluid solver based on a Gaussian spatial representation.
Compared to established grid-based fluid solvers, our framework offers the advantages of enhanced spatial details and effective vorticity preservation over time. When compared to continuous implicit representations, our method excels in handling boundary phenomena. Additionally, our approach is more stable, scalable, and efficient, supporting long time horizons with consistent parameters, making it a promising tool for both 2D and 3D fluid simulations.

Despite these strengths, our method has some limitations. First, relying on soft constraints to solve the Navier-Stokes equations introduces small residual errors in divergence and boundary conditions, which may lead to discrepancies in global fluid behavior when compared to grid-based solvers. Additionally, while harmonic components are implicitly preserved, they are not explicitly modeled with time consistency, which may result in inaccuracies when simulating dynamics in non-simply connected domains. Future work will focus on incorporating hard constraints and improving the modeling of harmonic components. Furthermore, we plan to explore inverse problems, such as key-frame-based fluid control, leveraging the efficient gradient calculation capabilities of our representation.

\begin{acks}
We thank the anonymous reviewers for their constructive comments.
\end{acks}

\bibliographystyle{ACM-Reference-Format}
\bibliography{reference}
\clearpage
\clearpage



\end{document}